\documentclass[12pt,a4paper]{article}

\usepackage{bm}

\usepackage{amsmath}
\usepackage{amsfonts}
\usepackage{amssymb}

\usepackage{colortbl}
\usepackage{hhline}

\newcommand{\sh}{\cellcolor[gray]{0.9}}

\definecolor{refkey}{rgb}{1,0,0}
\definecolor{labelkey}{rgb}{0,0,1}
\usepackage[pdftex]{graphicx}
\usepackage{tikz}
\usetikzlibrary{shapes,arrows,decorations.markings, intersections, calc}
\tikzset{
  ->-/.style={decoration={markings, mark=at position 0.55 with {\arrow[scale=1.5]{latex}}},
              postaction={decorate}},
  ->>-/.style={decoration={markings, mark=at position 0.47 with {\arrow[scale=1.5]{latex}}, mark=at position 0.71 with {\arrow[scale=1.5]{latex}}},
              postaction={decorate}},
  ->>>-/.style={decoration={markings, mark=at position 0.47 with {\arrow[scale=2]{latex}}, mark=at position 0.59 with {\arrow[scale=2]{latex}}, mark=at position 0.71 with {\arrow[scale=2]{latex}}}, postaction={decorate}}
}

\makeatletter
  
  \@addtoreset{equation}{section}
\makeatother

\newcommand{\CC}{\mathbb{C}}
\newcommand{\ZZ}{\mathbb{Z}}
\newcommand{\RR}{\mathbb{R}}

\newcommand{\ol}{\overline}
\newcommand{\wt}{\widetilde}

\def\tr{\mathop{\rm tr}\nolimits}

\def\diag{\mathop{\rm diag}\nolimits}
\def\Pexp{\mathop{\rm Pexp}\nolimits}

\def\Hom{\mathop{\rm Hom}\nolimits}

\newcommand{\sI}{\mathsf{I}}
\newcommand{\nn}{\nonumber}
\newcommand{\wh}{\widehat}

\begin{document}

%titlepage
\begin{titlepage}
\title{
\vspace{-1.5cm}
\begin{flushright}
{\normalsize TIT/HEP-674\\ July 2019}
\end{flushright}
\vspace{1.5cm}
\LARGE{Finite $N$ corrections to the superconformal index
of orbifold quiver gauge theories}}
\author{
Reona {\scshape Arai\footnote{E-mail: r.arai@th.phys.titech.ac.jp}},
Shota {\scshape Fujiwara\footnote{E-mail: s.fujiwara@th.phys.titech.ac.jp}},\\
Yosuke {\scshape Imamura\footnote{E-mail: imamura@phys.titech.ac.jp}},
and
Tatsuya {\scshape Mori\footnote{E-mail: t.mori@th.phys.titech.ac.jp}}
\\
\\
{\itshape Department of Physics, Tokyo Institute of Technology}, \\ {\itshape Tokyo 152-8551, Japan}}

\date{}
%\date{July 12, 2019}
\maketitle
\thispagestyle{empty}

%abstract
\begin{abstract}
We investigate the AdS/CFT correspondence for
quiver gauge theories realized on D3-branes
put on abelian orbifolds by using the superconformal index.
We assume that on the gravity side
the finite $N$ corrections of the index are
reproduced by D3-branes wrapped on three particular
three-cycles in the internal space ${\cal Y}$,
the abelian orbifold of $\bm{S}^5$.
We first establish the relation between
baryonic charges on the gauge theory side
and the D3-brane wrapping numbers and holonomies
on D3-branes.
Then we confirm our proposal
by comparing the results of localization
for gauge theories and the results on the AdS side
including the contributions of D3-branes and excitation on them
for many examples.
We only focus on the leading finite $N$ corrections
starting from $q^N$,
and leave the sub-leading corrections
starting at $q^{kN}$ ($k\geq2$) as a task for
the future.
We find complete agreement for the leading corrections
in all examples.
\end{abstract}
\end{titlepage}
\tableofcontents
%%%%%%%%%%%%%%%%%%%%%%%%%%%%%%%%%%%%%%%%%%%%%%%%%%%%%%%
\section{Introduction}

An ${\cal N}=1$ gauge theory
is realized on the worldvolume of $N$ D3-branes put
at the apex of a Calabi-Yau cone ${\cal X}$.
If $N$ is sufficiently large
the D3-brane system is well described as the classical supergravity solution
$AdS_5\times{\cal Y}$,
where ${\cal Y}$ is the base of the cone ${\cal X}$.
The AdS/CFT correspondence \cite{Maldacena:1997re,Witten:1998qj,Gubser:1998bc}
claims that the type IIB string theory in this background
is dual to the superconformal gauge theory
realized on the D3-branes at IR.
This duality has been tested by calculating various quantities
on the both sides and confirming their agreement.
In this paper we focus on the superconformal index \cite{Kinney:2005ej}.
We define the index as a formal power series of $q^{\frac{1}{2}}$ with
the coefficient
of each term being a Laurent polynomial of other fugacities.
A BPS operator with the dimension $d$ and the right-handed spin $\ol j$
contributes $\propto q^{d+\ol j}$ to the index.
See (\ref{indexdef0}) for an explicit definition.

In the large $N$ limit the agreement of the superconformal index
has been confirmed in a large class of the internal spaces ${\cal Y}$ and the
corresponding superconformal field theories.
For ${\cal N}=4$ $U(N)$ SYM corresponding to ${\cal Y}=\bm{S}^5$,
the large $N$ index was calculated on the both sides in \cite{Kinney:2005ej} and
agreement was confirmed.
The AdS/CFT correspondence for orbifolds $\bm{S}^5/\Gamma$ was suggested in
\cite{Kachru:1998ys,Lawrence:1998ja}.
The agreement of the index for the orbifolds $\bm{S}^5/\ZZ_n$ with an $A_{n-1}$ type fixed locus
$\bm{S}^1\subset {\cal Y}$
and the corresponding quiver gauge theories
was confirmed in \cite{Nakayama:2005mf}.
It was found that on the gravity side not only the gravity multiplet
but also the tensor multiplets living on the fixed locus contribute to
the index.
The index of the Kaluza-Klein modes in $AdS_5\times T^{1,1}$
was calculated in \cite{Nakayama:2006ur}.
The large $N$ index of the corresponding quiver gauge theory,
the Klebanov-Witten theory \cite{Klebanov:1998hh},
was calculated in \cite{Gadde:2010en}
and the agreement was confirmed.

If $N$ is finite the correspondence
is modified for operators with dimension of order $N$ or
larger.
For example, in ${\cal N}=4$ $SO(2N)$ SYM
the Pfaffian operators with dimension $N$ correspond
not to Kaluza-Klein modes of supergravity fields but to
D3-branes wrapped around topologically non-trivial cycles
in $\bm{S}^5/\ZZ_2$ \cite{Witten:1998xy}.
Similar relations hold for baryonic operators in quiver gauge theories
whose dual geometries have topologically non-trivial
three-cycles \cite{Gukov:1998kn}.

D3-branes play a role in finite $N$ corrections
even when the internal space does not have
topologically non-trivial three-cycles.
In the case of ${\cal N}=4$ $U(N)$ SYM
the one-to-one correspondence between BPS
operators and Kaluza-Klein modes
is broken down for operators with dimension
of order $N$ or larger.
On the gauge theory side this is because
single trace operators with the length $L>N$ are
not independent but decomposable into
shorter single-trace operators.
We can explain this on the gravity side by assuming
that the operators correspond not to point-like gravitons
but to giant gravitons:
D3-branes wrapped on topologically trivial three-cycles in $\bm{S}^5$.
$1/2$ BPS giant gravitons were constructed in \cite{McGreevy:2000cw}.
Their angular momentum $J$ in $\bm{S}^5$ has the
upper bound $J\leq N$, and
the absence of giant gravitons with $J>N$
corresponds to the absence of independent single-trace operators
with the length $L>N$.
Indeed, the BPS partition function \cite{Benvenuti:2006qr}
of ${\cal N}=4$ $U(N)$ SYM with
finite $N$ was exactly reproduced in \cite{Biswas:2006tj}
based on the idea of \cite{Beasley:2002xv}
by the geometric quantization of $1/8$ BPS giant gravitons
constructed in \cite{Mikhailov:2000ya}.
There is also a complementary way to reproduce the
same BPS partition function \cite{Mandal:2006tk}
by using giant gravitons expanded in $AdS_5$ \cite{Grisaru:2000zn,Hashimoto:2000zp}.

The result in \cite{Beasley:2002xv} was extended in \cite{Arai:2018utu}
to S-fold theories including
the ${\cal N}=4$ $SO(2N)$ SYM.
The exact BPS partition function was
derived by
the geometric quantization of BPS configurations
of D3-branes in $\bm{S}^5/\ZZ_k$,
where $\ZZ_k$ is the S-fold action
transforming the $(p,q)$-string charges non-trivially.
It is natural to attempt a similar derivation for the
superconformal index, and indeed in \cite{Arai:2019xmp}
it was shown for the S-fold theories
that the leading finite $N$ corrections can be reproduced as the
index of fluctuation modes on D3-branes wrapped around
particular three-cycles.
We mean in this paper by ``the leading finite $N$ corrections''
the corrections starting from ${\cal O}(q^N)$.
In the case of S-fold theories
we also have ``the sub-leading corrections''
starting from ${\cal O}(q^{2N})$,
which was not studied in \cite{Arai:2019xmp}.

The purpose of this paper is to calculate
the finite $N$ corrections to the superconformal
index for ordinary orbifolds ${\cal Y}={\bm S}^5/\wt\Gamma$,
which have much more variety
than the S-folds.
(We use $\wt\Gamma$ for the orbifold group rather than $\Gamma$ and $\Gamma$ for the dual group
$\Hom(\wt\Gamma,U(1))$ because in the following we use the dual group more frequently
than the orbifold group itself.)
For an ${\cal N}=1$ supersymmetry to be preserved
$\wt\Gamma$ must be a finite subgroup of $SU(3)$ acting on
$\CC^3$ coordinates $(X,Y,Z)$
\footnote{We also use the notation $X_I$ which means $X,Y,$ and $Z$ for $I=X,Y,$ and $Z$.}.
We restrict our attention to the toric case.
Namely, we assume that $\wt\Gamma$ is abelian,
and is a subset of the Cartan subgroup $H=U(1)^2\subset SU(3)$.

It is convenient to represent $h\in H$ as the
$3\times 3$ matrix acting on $(X,Y,Z)$:
\begin{align}
w_X(h)^{R_X}
w_Y(h)^{R_Y}
w_Z(h)^{R_Z}
=
\left(\begin{array}{ccc}
w_X(h) \\
& w_Y(h) \\
&& w_Z(h)
\end{array}\right).
\label{c3action}
\end{align}
$R_X$, $R_Y$, and $R_Z$ are
the Cartan generators of $su(4)\sim so(6)$
acting on $X$, $Y$, and $Z$, respectively,
and $w_I(h)$ ($I=X,Y,Z$) are complex numbers with the absolute value $1$
depending on $h\in H$.
For this to be an element of $SU(3)$
$w_I(h)$ must satisfy
\begin{align}
w_X(h)w_Y(h)w_Z(h)=1\quad
\forall h\in H.
\label{speciality}
\end{align}

Generically the orbifolding breaks the ${\cal N}=4$ supersymmetry
down to ${\cal N}=1$.
Let $Q_a$ and $\ol Q^{\dot a}$ be the left-handed and the right-handed unbroken supercharges, respectively.
To define the superconformal index
we use $\ol Q^{\dot 1}$
such that
$\ol\Delta=2\{(\ol Q^{\dot 1})^\dagger,\ol Q^{\dot 1}\}$
is given by
\begin{align}
\ol\Delta=2\{(\ol Q^{\dot 1})^\dagger,\ol Q^{\dot 1}\}=H-2\ol J-(R_X+R_Y+R_Z).
\end{align}
The index is defined by\footnote{%
Our convention and notation for the fugacities are the same
as those in \cite{Arai:2019xmp} except for $y$,
which is denoted in \cite{Arai:2019xmp} by $\wt y$.}
\begin{align}
{\cal I}(q,y,u_I,\zeta)=\tr
\left[(-1)^F\ol x^{\ol\Delta}q^{H+\ol J}
y^{2J}
\zeta^{\bm b}
\prod_{I=X,Y,Z}u_I^{R_I}
\right].
\label{indexdef0}
\end{align}
Only BPS operators saturating the BPS bound
$\ol\Delta\geq0$
contribute to the index and hence the index
is independent of $\ol x$.
$H$ is the dilatation and
$J$ and $\ol J$ are the left- and right-angular momenta
normalized so that the eigenvalues are
quantized with unit $1/2$.
$u_I$ are $SU(3)$ fugacities satisfying $u_Xu_Yu_Z=1$.
We also use two independent variables $u$ and $v$ related to $u_I$ by
$(u_X,u_Y,u_Z)=(u,\frac{v}{u},\frac{1}{v})$.
$\bm{b}$ and $\zeta$ collectively represent
baryonic charges and the corresponding fugacities,
respectively,
which will be discussed later in detail.

From the viewpoint on the gravity side
this index is expected to be factorized into two factors:
\begin{align}
{\cal I}={\cal I}^{\rm KK}
{\cal I}^{\rm D3}.
\end{align}
${\cal I}^{\rm KK}$
is the contribution of Kaluza-Klein modes
of massless fields, which has been already studied in the literature.
${\cal I}^{\rm KK}$ gives
the exact index in the large $N$ limit.
The purpose of this paper is to investigate the
other factor, ${\cal I}^{\rm D3}$,
which gives finite $N$ corrections due to
D3-branes wrapped on three-cycles in ${\cal Y}$.
Again we focus only on the leading corrections starting from
${\cal O}(q^N)$,
and do not pay attention to the sub-leading corrections
starting from ${\cal O}(q^{kN})$ with $k\geq2$ depending on
the sector we consider.

The rest of this paper is organized as follows.
In the next section we summarize the toric diagrams and the quiver gauge
theories for abelian orbifolds ${\cal Y}=\bm{S}^5/\wt\Gamma$.
In section \ref{largen.sec}
we review how the superconformal index in the
large $N$ limit is calculated on the gravity side
as the contribution of Kaluza-Klein modes in the orbifold ${\cal Y}=\bm{S}^5/\wt\Gamma$.
In section \ref{finiten.sec} we explain a prescription
to obtain the leading finite $N$ corrections
from wrapped D3-branes
based on the analysis in \cite{Arai:2019xmp}.
We first discuss the relation between
the wrapping number of D3-branes and the baryonic charges in the
quiver gauge theory,
and then we give a prescription
to calculate the contribution
in each wrapping sector.
In section \ref{examples.sec}
we apply the method to examples with $N=2$
and confirm the agreement
between the results of the localization
and those of the D3-brane analysis.
Section \ref{discussion.sec} is devoted to
summary and discussions.
The appendix contains results for $N=3$ for some of
orbifolds in section \ref{examples.sec}.

To write down indices we use characters associated with
the spin and flavor symmetries.
They are defined as follows.

The spin characters $\chi^J_n$ are defined by
\begin{align}
\chi^J_n=\frac{y^{n+1}-y^{-(n+1)}}{y-y^{-1}}.
\end{align}

The $u(2)$ characters $\chi_n(a,b)$ are defined by
\begin{align}
\chi_n(a,b)=\frac{a^{n+1}-b^{n+1}}{a-b}.\label{u2charactor}
\end{align}
These are used to give the index of a theory with an $SU(2)$ flavor symmetry.
In all examples in this paper the $SU(2)$ acts on $Y$ and $Z$,
and the arguments of the characters are $u_Y=\frac{v}{u}$ and $u_Z=\frac{1}{v}$.
We use the short-hand notation $\chi_n=\chi_n(\frac{v}{u},\frac{1}{v})$.

$\chi_{(r_1,r_2)}$ are the $su(3)$ characters for representations with the Dynkin labels
$(r_1,r_2)$.
For the fundamental and the anti-fundamental representations
these are given by
\begin{align}
\chi_{(1,0)}=\sum_{I=X,Y,Z}u_I,\quad
\chi_{(0,1)}=\sum_{I=X,Y,Z}u_I^{-1}.
\end{align}
For a general representation $(r_1,r_2)$ it is given by
\begin{align}
\chi_{(r_1,r_2)}=
\left|\begin{array}{ccc}
u_X^{r_1+1} & 1 & u_X^{-(r_2+1)} \\
u_Y^{r_1+1} & 1 & u_Y^{-(r_2+1)} \\
u_Z^{r_1+1} & 1 & u_Z^{-(r_2+1)}
\end{array}\right|
\Bigg/
\left|\begin{array}{ccc}
u_X & 1 & u_X^{-1} \\
u_Y & 1 & u_Y^{-1} \\
u_Z & 1 & u_Z^{-1}
\end{array}\right|.
\end{align}

For later use we define $\omega_n$ by
\begin{align}
\omega_n=\exp\left(\frac{2\pi i}{n}\right).
\end{align}

%%%%%%%%%%%%%%%%%%%%%%%%%%%%%%%%%%%%%%%%%%%%%%%%%%
\section{Orbifolds and quiver gauge theories}\label{orbifolds.sec}
\subsection{Toric diagrams}

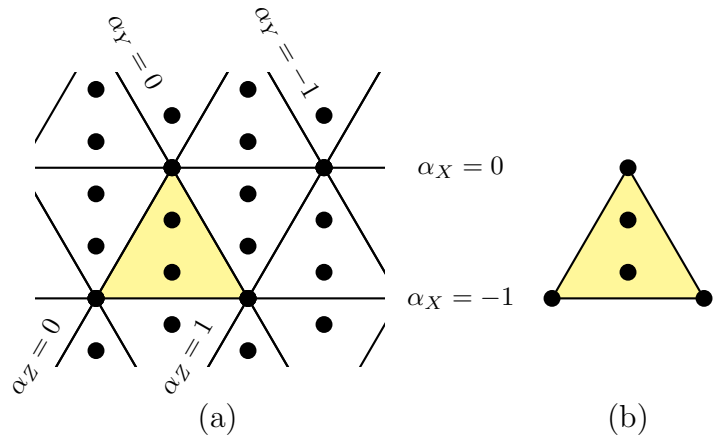
\begin{figure}[htb]
\centering
\begin{tikzpicture}
%Line label
\node [rotate=-60] at (-1.5,{2*sqrt(3)-0.2}) {\footnotesize{$\alpha_Y=0$}};
\node [rotate=-60] at (0.5,{2*sqrt(3)-0.2}) {\footnotesize{$\alpha_Y=-1$}};
\node [] at (2.8,0) {\footnotesize{$\alpha_X=-1$}};
\node [] at (2.8,{sqrt(3)}) {\footnotesize{$\alpha_X=0$}};
\node [rotate=60] at (-2.8,{-sqrt(3)/3-0.2}) {\footnotesize{$\alpha_Z=0$}};
\node [rotate=60] at (-0.8,{-sqrt(3)/3-0.2}) {\footnotesize{$\alpha_Z=1$}};

\node [] at (-0.4,-1.6) {(a)};
\node [] at (5,-1.6) {(b)};

%(b)
\fill [yellow!50] (6,0) -- (5,{sqrt(3)}) -- (4,0);
    \node [draw=black, circle, scale=0.5, thick, fill=black] (toric1) at (6,0) {};
    \node [draw=black, circle, scale=0.5, thick, fill=black] (toric2) at ($(toric1)+(-1,{sqrt(3)})$) {};
    \node [draw=black, circle, scale=0.5, thick, fill=black] (toric3) at ($(toric1)+(-2,0)$) {};
    \node [draw=black, circle, scale=0.5, thick, fill=black] (toric4) at ($(toric1)+(-1,{3/5*sqrt(3)})$) {};
    \node [draw=black, circle, scale=0.5, thick, fill=black] (toric5) at ($(toric1)+(-1,{1/5*sqrt(3)})$) {};
    \draw [thick](toric1) -- (toric2);
    \draw [thick](toric2) -- (toric3);
    \draw [thick](toric3) -- (toric1);

%(a)
\clip (-2.8,-0.9) rectangle (1.8,3);
\fill [yellow!50] (0,0) -- (-1,{sqrt(3)}) -- (-2,0);
\foreach \x / \y in {-2/0, 0/0, 2/0, 1/sqrt(3), -1/sqrt(3), 3/sqrt(3), 0/2*sqrt(3), 2/2*sqrt(3), -1/-sqrt(3), 1/-sqrt(3)}
{
    \node [draw=black, circle, scale=0.5, thick, fill=black] (toric1) at ({\x},{\y}) {};
    \node [draw=black, circle, scale=0.5, thick, fill=black] (toric2) at ($(toric1)+(-1,{sqrt(3)})$) {};
    \node [draw=black, circle, scale=0.5, thick, fill=black] (toric3) at ($(toric1)+(-2,0)$) {};
    \node [draw=black, circle, scale=0.5, thick, fill=black] (toric4) at ($(toric1)+(-1,{3/5*sqrt(3)})$) {};
    \node [draw=black, circle, scale=0.5, thick, fill=black] (toric5) at ($(toric1)+(-1,{1/5*sqrt(3)})$) {};
    \node [draw=black, circle, scale=0.5, thick, fill=black] (toric6) at ($(toric1)+(-1,{-sqrt(3)})$) {};
    \node [draw=black, circle, scale=0.5, thick, fill=black] (toric7) at ($(toric1)+(-1,{-3/5*sqrt(3)})$) {};
    \node [draw=black, circle, scale=0.5, thick, fill=black] (toric8) at ($(toric1)+(-1,{-1/5*sqrt(3)})$) {};
    \draw [thick](toric1) -- (toric2);
    \draw [thick](toric2) -- (toric3);
    \draw [thick](toric3) -- (toric1);
    \draw [thick](toric3) -- (toric6);
    \draw [thick](toric6) -- (toric1);
}
\end{tikzpicture}
\caption{(a) The $\ol H$ plane with the lattice $P$ for the orbifold group
$\wt\Gamma=\ZZ_5$ generated by $\diag(\omega_5^{-2},\omega_5,\omega_5)$ is shown.
$L$ is expressed as the set of
intersections of three sets of parallel lines.
The lattice $P$ is shown by dots.
(b)
A triangle picked up from (a) is shown.
This is nothing but the toric diagram of the orbifold
$\CC^3/\ZZ_5$.}\label{wth.eps}
\end{figure}

Orbifolds we analyze in this paper are
a special class of toric Calabi-Yaus,
and their structure can be expressed
by using toric diagrams.
To define the toric diagram
it is convenient to define $\alpha_I$ ($I=X,Y,Z$) by
\begin{align}
w_I=e^{2\pi i\alpha_I},\quad
\alpha_X+\alpha_Y+\alpha_Z=0.
\end{align}
The parameters
$\alpha_I$ can be regarded as
(redundant) coordinates of the covering space $\ol H=\RR^2$ of $H$.
These are angular variables with the period $1$.
Let $L$ be the associated lattice defined by $\alpha_I\in \ZZ$.
$H$ and $\ol H$ are related by $H=\ol H/L$.
In Figure \ref{wth.eps} (a) the lattice $L$ is expressed as
the set of intersections of three sets of parallel
lines $\alpha_I\in\ZZ$,
which give a tessellation of the plane $\ol H$ by congruent triangles.

$\wt\Gamma$ is a finite subgroup of $H$, and is expressed as
a lattice $P$ in $\ol H$, which includes $L$ as a sublattice.
In other words, $P$ is a refinement of $L$.
Thanks to the periodicity and the $\ZZ_2$ rotational symmetry of the lattice $P$
all triangles contain points in $P$ in the same way,
and we can pick up one triangle to represent the orbifold group
$\wt\Gamma$ (See (b) in Figure \ref{wth.eps}).
This is nothing but the toric diagram of the orbifold.%
\footnote{Usually a toric diagram is drawn so that $P$ is a square lattice.
We do not do so and we draw the
diagram as a regular triangle with dots.}

Note that
if we express $\wt g\in\wt\Gamma$
in the matrix form (\ref{c3action})
each of diagonal components $w_I(\wt g)$ is
a one-dimensional representation of $\wt\Gamma$.
Namely, $w_I$ can be regarded as elements of the
dual group
\begin{align}
\Gamma=\Hom(\wt\Gamma,U(1)).
\label{dualgroup}
\end{align}
Furthermore, the group $\Gamma$ is generated by the elements $w_I$.
We can specify the orbifold ${\cal X}=\CC^3/\wt\Gamma$ by giving
a set of relations satisfied by $w_I$,
which must always include $w_Xw_Yw_Z=e$, where $e$ is the identity element of $\Gamma$.

%%%%%%%%%%%%%%%%%%%%%%%%%%%%%%%%%%%%%%%%%%%%%%%%%%%
\subsection{Quiver gauge theories}\label{quivergt.sec}
The quiver gauge theory realized on an orbifold ${\cal X}=\CC^3/\wt\Gamma$
is obtained by the standard prescription
\cite{Douglas:1996sw,Douglas:1997de} as follows.
We start from ${\cal N}=4$ SYM with the gauge group $U(|\Gamma|N)$,
where $|\Gamma|$ is the order of $\Gamma$, which is the same as the order of $\wt\Gamma$.
The ${\cal N}=4$ vector multiplet  consists of
an ${\cal N}=1$ vector multiplet $V$ and three
${\cal N}=1$ chiral multiplets $\Phi_I$ ($I=X,Y,Z$).
Because we consider abelian orbifolds we
can discuss the projection on each ${\cal N}=1$ multiplet separately.

For the vector multiplet $V$ the action of $\wt g\in\wt\Gamma$ is
\begin{align}
\wt g:V\rightarrow V'=u(\wt g)Vu^{-1}(\wt g),
\label{holonomy}
\end{align}
where
$u\in \Hom(\wt\Gamma,U(|\Gamma|N))$
represents the action associated with a holonomy.
In this paper we consider only the case that
all gauge groups have the same rank,
and it is realized by taking
\begin{align}
u(\wt g)=R(\wt g)\otimes\bm{1}_N,
\end{align}
where $R(\wt g)$ is the regular representation of $\wt\Gamma$.
The regular representation of a finite group is the direct sum
of all irreducible representations.
Because $\wt\Gamma$ is abelian irreducible representations are
identified with elements of the dual group (\ref{dualgroup})
and
the regular representation is
given by
\begin{align}
R(\wt g)=\bigoplus_{g\in\Gamma}g(\wt g).
\label{regularrep}
\end{align}
Correspondingly, we divide $|\Gamma|N\times|\Gamma|N$ matrix $V$
into $|\Gamma|^2$ blocks $V_{g_1g_2}$ of size $N\times N$ labeled
by $g_1,g_2\in\Gamma$.
The action of $\wt g\in\wt\Gamma$ on each block is
\begin{align}
\wt g:V_{g_1g_2}\rightarrow
V'_{g_1g_2}=
\frac{g_1(\wt g)}{g_2(\wt g)}V_{g_1g_2}.
\end{align}
Then the projection leaves diagonal blocks $V_{gg}$.
Let $U(N)_g=SU(N)_g\times U(1)_g$ be the symmetry corresponding to the block $V_{gg}$.
We define
\begin{align}
G=\prod_{g\in\Gamma}SU(N)_g,\quad
G_B^0=\prod_{g\in\Gamma}U(1)_g.
\label{gaugegroup}
\end{align}
$G$ is the gauge group.
The gauge fields for $G_B^0$ are decoupled in the IR
and $G_B^0$ becomes global symmetry.
In general $G_B^0$ is broken by anomalies to a subgroup
$G_B\subset G_B^0$.

For the chiral multiplets the orbifold action
is the composition of the $\CC^3$ rotation
given by (\ref{c3action}) and the holonomy action like
(\ref{holonomy}).
Corresponding to the irreducible decomposition
(\ref{regularrep}) the chiral multiplets are also
decomposed into blocks $(\Phi_I)_{g_1g_2}$ of size $N\times N$.
$(\Phi_I)_{g_1g_2}$ belongs to the bi-fundamental representation $(N,\ol N)$
of $SU(N)_{g_1}\times SU(N)_{g_2}$, and in the following we use the notation
$\Phi_I^{g_1\rightarrow g_2}$.
The $\wt g\in\wt\Gamma$ action on each block is
\begin{align}
\wt g:\Phi_I^{g_1\rightarrow g_2}\rightarrow
\Phi'^{g_1\rightarrow g_2}_I=
w_I(\wt g)\frac{g_1(\wt g)}{g_2(\wt g)}\Phi_I^{g_1\rightarrow g_2},
\end{align}
and this block remains after the orbifold projection
if and only if the relation
\begin{align}
w_Ig_1=g_2
\label{remaining}
\end{align}
holds\footnote{In this relation $w_I$, $g_1$, and $g_2$ are elements of $\Gamma$.
Namely, this relation means that $w_I(\wt g)g_1(\wt g)=g_2(\wt g)$ holds for all $\wt g\in\wt\Gamma$.}.
This condition determines the matter contents of the
quiver gauge theory.
Namely, we are left with $3|\Gamma|$ bi-fundamental
fields $\Phi_I^{g\rightarrow w_Ig}$ labeled by
$I=X,Y,Z$ and $g\in\Gamma$.
With this information it is easy to draw the quiver diagram.
We first draw $|\Gamma|$ vertices corresponding to the elements of
$\Gamma$.
Each of them represents an $SU(N)$ gauge group.
A chiral multiplet
belonging to the $(N,\ol N)$ representation
of $SU(N)_{g_1}\times SU(N)_{g_2}$
is represented
as an arrow from the vertex $g_1$ to $g_2$.
Because $\Gamma$ is a finite abelian group and
can be given as a subgroup of $U(1)^2$
it is natural and convenient to draw the diagram on the
torus, and such a diagram is called
a periodic quiver diagram.

Once we obtain the field contents
of the quiver gauge theory we can in principle
calculate the superconformal index
for an arbitrary finite $N$ by the localization formula
\begin{align}
\mathcal{I}
=\int d\mu\Pexp\left(\sum_{g\in\Gamma}\left(
\sI_v^g
+\sI_X^{g\rightarrow w_Xg}
+\sI_Y^{g\rightarrow w_Yg}
+\sI_Z^{g\rightarrow w_Zg}
\right)\right).
\label{indexformula}
\end{align}
The plethystic exponential $\Pexp$ is defined as
\begin{align}
\Pexp (f(x_i))=\exp \left (\sum _{n=1}^{\infty }\frac{1}{n}f(x_i^n)\right ).
\end{align}
$\sI_{v}^g$ is the single particle index of the $SU(N)_g$ vector multiplet:
\begin{align}
\sI_{v}^g=\left(
-\dfrac{yq^{\frac{3}{2}}}{1-yq^{\frac{3}{2}}}
-\dfrac{y^{-1}q^{\frac{3}{2}}}{1-y^{-1}q^{\frac{3}{2}}}\right)
\chi_{\mathrm{adj}}^g,
\label{suni}
\end{align}
where $\chi_{\rm adj}^g$ is the
character of the adjoint representation of $SU(N)_g$.
The single-particle index of the chiral multiplet
$\Phi_I^{g_1\rightarrow g_2}$ is
\begin{align}
\sI_{I}^{g_1\rightarrow g_2}
=\dfrac{q u_I\chi^{g_1}_N\chi^{g_2}_{\overline{N}}\zeta_{g_1}^{\frac{1}{N}}\zeta_{g_2}^{-\frac{1}{N}}
-q^2 u_I^{-1}\chi^{g_2}_N {\chi}^{g_1}_{\overline{N}}\zeta_{g_1}^{-\frac{1}{N}}\zeta_{g_2}^{\frac{1}{N}}
}{(1-yq^{\frac{3}{2}})(1-y^{-1}q^{\frac{3}{2}})},
\label{spixyz}
\end{align}
where
$\chi_N^g$ and $\ol\chi_N^g$ are the characters of the
fundamental and the anti-fundamental representations of
$SU(N)_g$.
The explicit forms of the characters are
\begin{align}
\chi_N^g=\sum_{a=1}^N z_a^g,\quad
\chi_{\ol N}^g=\sum_{a=1}^N \frac{1}{z_a^g},\quad
\chi_{\rm adj}^g=\chi_N^g\chi_{\ol N}^g-1.
\end{align}
Because the gauge groups are $SU(N)$ the gauge fugacities $z_a^g$ ($a=1,\ldots,N$)
are constrained by
\begin{align}
\prod_{a=1}^Nz_a^g=1.
\end{align}
$\zeta_g$ are fugacities of $U(1)_g$ baryonic symmetries.
We denote the set of the $|\Gamma|$ fugacities collectively by $\zeta$
and it can be regarded as an element of the classical baryonic symmetry $G_B^0$.
If the baryonic symmetry is anomalous and is broken to its subgroup,
$\zeta$ takes only values in the
anomaly-free subgroup.

$\int d\mu$ is the integration over the gauge fugacities $z_a^g$
defined by
\begin{align}
\int d\mu =\prod_{g\in\Gamma}\frac{1}{N!}\oint\left(\prod_{a=1}^{N-1}\frac{dz_a^g}{2\pi iz_a^g}\right)
\prod_{a\neq b}(1-\frac{z_a^g}{z_b^g}).
\end{align}

Although the single-particle indices (\ref{spixyz})
include fractional powers of $\zeta_g$
only integral powers remain after the integration over the gauge fugacities.

%%%%%%%%%%%%%%%%%%%%%%%%%%%%%%%%%%%%%%%%%%%%%%%%%%%%%
\subsection{Baryonic charges}
Let $\Lambda_B^0$ be the charge lattice associated with the classical baryonic symmetry $G_B^0$.
An element of $\Lambda_B^0$ is specified by the set of $|\Gamma|$ $U(1)_g$ charges $B_g\in\ZZ$
satisfying
\begin{align}
\sum_{g\in\Gamma}B_g=0,
\label{total0}
\end{align}
because all matter fields belong to bi-fundamental representations%
\footnote{We normalize $B_g$ so that a field belonging to the $SU(N)_g$ fundamental (anti-fundamental) representation
carries $B_g=+1/N$ ($-1/N$).}.
Let $\bm{b}_g$ be the formal basis vector of the $U(1)_g$ charge.
A general element of $\Lambda_B^0$ is given by
\begin{align}
\bm{b}=\sum_{g\in\Gamma}B_g\bm{b}_g.
\label{vectorb}
\end{align}
Due to (\ref{total0})
the lattice $\Lambda_B^0$ is
spanned by $|\Gamma|-1$ vectors $\bm{b}_e-\bm{b}_g$ ($g\neq e$).

Now let us consider the effect of anomalies.
If the theory is chiral, the presence
of instantons causes violation
of the conservation laws of the charges $B_g$.
Let $N_g$ be the instanton numbers of $SU(N)_g$.
Then the baryonic charges change in this instanton background by
\begin{align}
\Delta B_g=\sum_{g'\in\Gamma}n_{g\rightarrow g'}N_{g'},
\label{deltab}
\end{align}
where $n_{g\rightarrow g'}$ is the number of the arrows in the quiver diagram
from $g$ to $g'$.
Arrows in the opposite direction are counted by $-1$.
A single $SU(N)_g$ instanton
changes the vector $(\ref{vectorb})$ by
\begin{align}
\Delta\bm{b}_g\equiv
\sum_{g'}\bm{b}_{g'}n_{g'\rightarrow g}.
\end{align}
For the orbifold quiver gauge theories
this is given by
\begin{align}
\Delta\bm{b}_g=\sum_{I=X,Y,Z}
(\bm{b}_{w_I^{-1}g}
-\bm{b}_{w_Ig}).
\end{align}
The quantum baryonic charge lattice $\Lambda_B$ taking account of this breaking
of the conservation laws
is given by
\begin{align}
\Lambda_B=\Lambda_B^0/\stackrel{A}{\sim},
\end{align}
where $\stackrel{A}{\sim}$ is the equivalence relation defined by
\begin{align}
\Delta\bm{b}_g\stackrel{A}{\sim} 0\quad
\forall g\in\Gamma.
\label{anomaly}
\end{align}

$\zeta^{\bm{b}}$ in the definition of the index (\ref{indexdef0})
is a short-hand notation for $\prod_{g\in\Gamma}\zeta_g^{B_g}$.
If we treat fugacities $\zeta_g$
as independent variables among them
we can extract the index ${\cal I}_{\bm{b}}$
for each $\bm{b}$ by the expansion
\begin{align}
{\cal I}=\sum_{\bm{b}\in\Lambda_B^0}\zeta^{\bm{b}}{\cal I}_{\bm{b}}.
\label{clexpansion}
\end{align}
However, the physical meaning of ${\cal I}_{\bm{b}}$
is not clear because the classical
baryonic charges are not conserved
due to the anomalies.
Instead of defining the index for each $\bm{b}$
we should treat equivalence classes $\wh{\bm{b}}\in\Lambda_B$ as conserved charges
and define the index for
each of such classes by
\begin{align}
{\cal I}_{\wh{\bm{b}}}=\sum_{\bm{b}\in\wh{\bm{b}}}{\cal I}_{\bm{b}}.
\end{align}
The index for a class $\wh{\bm{b}}$ can be directly extracted
from the index (\ref{indexdef0}) by an expansion similar to
(\ref{clexpansion}).
We impose the conditions
\begin{align}
\zeta^{\Delta\bm{b}_g}=1\quad
\forall g\in\Gamma
\end{align}
on the fugacities $\zeta$.
This is equivalent to the requirement that
$\zeta$ is an element of the anomaly free
baryonic symmetry group $G_B\subset G_B^0$.
Then $\zeta^{\bm b}$ depends on
$\bm{b}$ only through the equivalence class
to which $\bm{b}$ belongs
and we denote it by $\zeta^{\wh{\bm{b}}}$.
We can rewrite the expansion (\ref{clexpansion})
as
\begin{align}
{\cal I}=\sum_{\wh{\bm{b}}\in\Lambda_B}\zeta^{\wh{\bm{b}}}{\cal I}_{\wh{\bm{b}}}.
\end{align}

%%%%%%%%%%%%%%%%%%%%%%%%%%%%%%%%%%%%%%%%%%%%%%%%%%%%%%%%%%%%%%%%%%%%%%%%%%%%
\section{Large $N$ limit}\label{largen.sec}
In this section we review known results for the superconformal index in the large $N$ limit.
As is mentioned in the introduction
it is given on the AdS side by ${\cal I}^{\rm KK}$,
the contribution of Kaluza-Klein modes of massless fields.
We can divide this into two factors
${\cal I}^{\rm grav}$
and ${\cal I}^{\rm tensor}$:
the index of the gravity multiplet in the bulk
and the index of the tensor multiplets
localized on the fixed loci.

%%%%%%%%%%%%%%%%%%%%%%%%%%%%%%%%%%%%%%%%%%%%%
\subsection{Gravity multiplet}
Let us first consider the gravity multiplet contribution ${\cal I}^{\rm grav}$.
It is given by
\begin{align}
{\cal I}^{\rm grav}=\Pexp\left({\cal P}\sI^{\rm grav}\right),
\label{igravdef}
\end{align}
where ${\cal P}$ is the orbifold projection operator
and $\sI^{\rm grav}$
is the single-particle index of the Kaluza-Klein modes in $AdS_5\times\bm{S}^5$.
For $\bm{S}^5$ without orbifolding
the modes belong to the series of ${\cal N}=4$ superconformal multiplets
${\cal B}^{\frac{1}{2},\frac{1}{2}}_{[0,n,0](0,0)}$ ($n=1,2,\ldots$)
\cite{Kim:1985ez,Gunaydin:1984fk}.
(See \cite{Dolan:2002zh} for the notation for the superconformal representations.)
The corresponding index
is given by \cite{Kinney:2005ej}
\begin{align}
\sum_{n=1}^\infty\sI_{{\cal B}^{\frac{1}{2},\frac{1}{2}}_{[0,n,0](0,0)}}
=\frac{uq}{1-uq}
+\frac{\frac{v}{u}q}{1-\frac{v}{u}q}
+\frac{\frac{1}{v}q}{1-\frac{1}{v}q}
-\frac{yq^{\frac{3}{2}}}{1-yq^{\frac{3}{2}}}
-\frac{\frac{1}{y}q^{\frac{3}{2}}}{1-\frac{1}{y}q^{\frac{3}{2}}}.
\end{align}
The same index is obtained as the large $N$ limit of the index of
${\cal N}=4$ SYM with the gauge group $U(N)$.
Because we consider quiver gauge theories with
the gauge group $SU(N)^{|\Gamma|}$,
we define $\sI^{\rm grav}$ by
subtracting the contribution of the ${\cal N}=1$ vector multiplet
\begin{align}
\sI_{\rm vec}=
-\frac{yq^{\frac{3}{2}}}{1-yq^{\frac{3}{2}}}
-\frac{\frac{1}{y}q^{\frac{3}{2}}}{1-\frac{1}{y}q^{\frac{3}{2}}}.
\label{n1vector}
\end{align}
Namely, we define $\sI^{\rm grav}$ by
\begin{align}
\sI^{\rm grav}
=\frac{uq}{1-uq}
+\frac{\frac{v}{u}q}{1-\frac{v}{u}q}
+\frac{\frac{1}{v}q}{1-\frac{1}{v}q}
=\sum_{I=1}^3\frac{u_Iq}{1-u_Iq}.
\label{igravdea}
\end{align}
We define the projection operator ${\cal P}$
so that it picks up the $\wt\Gamma$ invariant terms from
the index.
Let $f$ be a single-particle index defined by summing up modes
in $\bm{S}^5$:
\begin{align}
f(q,y,u_I)=\tr_{\bm{S}^5}\left(\cdots\prod_{I=X,Y,Z} u_I^{R_I}\right).
\end{align}
We here focus on the dependence on the $SU(3)$ fugacities $u_I$.
Then the projected index ${\cal P}f$ is given by
\begin{align}
{\cal P}f(q,y,u_I)
&=\tr_{\bm{S}^5}\left(\frac{1}{|\Gamma|}\sum_{\wt g\in\wt \Gamma}\wt g\cdots
\prod_{I=X,Y,Z}u_I^{R_I}
\right).
\nonumber\\
&=\frac{1}{|\Gamma|}\sum_{\wt g\in\wt\Gamma}\tr_{\bm{S}^5}\left(\cdots
\prod_{I=X,Y,Z}(w_I(\wt g)u_I)^{R_I}\right)
\nonumber\\
&=\frac{1}{|\Gamma|}\sum_{\wt g\in\wt\Gamma}f(q,y,w_I(\wt g)u_I).
\label{pgammadef}
\end{align}
At the second equality we used $\wt g=\prod_{I=X,Y,Z}w_I(\wt g)^{R_I}$.

%%%%%%%%%%%%%%%%%%%%%%%%%%%%%%%%%%%%%%%%%%%%%%%%%%%%
\subsection{Tensor multiplet}
The tensor multiplet contribution ${\cal I}^{\rm tensor}$ is present when the
orbifold has fixed loci \cite{Nakayama:2005mf}.

The presence of fixed loci and their types are
easily read off from the toric diagram.
The orbifold ${\cal Y}=\bm{S}^5/\wt\Gamma$ has
fixed points if there exist $\wt g\in\wt\Gamma$ such that $\wt g\neq\wt e$ and
$w_I(\wt g)=1$ for one of $I=X,Y,Z$.
In the toric diagram such an element is expressed as a dot on an open edge.
(By an edge side we mean an edge with two endpoints excluded.)
For example, if there are $k-1$ dots on an open edge
and the edge is divided by
them into $k$ pieces,
then the $k-1$ elements of $\wt\Gamma$ associated with the dots together with
the identity element
form $\ZZ_k\subset\wt\Gamma$
acting only on two coordinates, and the fixed locus
is the set of $A_{k-1}$ type singularities.

Let us consider the orbifold
${\cal X}=\CC\times\CC^2/\ZZ_n$
with the orbifold group generated by
$\diag(1,\omega_n,\omega_n^{-1})$.
In this case an ${\cal N}=2$ supersymmetry is preserved.
This has the $A_{n-1}$-type fixed locus $\bm{S}^1\subset{\cal Y}$ given by $Y=Z=0$.
In the six-dimensional space $AdS_5\times\bm{S}^1$ the $A_{n-1}$ type ${\cal N}=(2,0)$ theory
lives, and $n-1$ tensor multiplets in the theory contribute to the index \cite{Nakayama:2005mf}.
The Kaluza-Klein modes of a single tensor multiplet belong
to the series of ${\cal N}=2$ superconformal representations
${\cal E}_{m(0,0)}$ ($m=1,2,3,\ldots$), and the corresponding single-particle index is
\begin{align}
\sum_{m=1}^\infty\sI_{{\cal E}_{m(0,0)}}
&=\frac{uq}{1-uq}
-\frac{yq^{\frac{3}{2}}}{1-yq^{\frac{3}{2}}}
-\frac{\frac{1}{y}q^{\frac{3}{2}}}{1-\frac{1}{y}q^{\frac{3}{2}}}.
\end{align}
(We can easily calculate this index by using the method in \cite{Cordova:2016xhm}.)
Similarly to the gravity multiplet contribution,
we subtract the contribution of the ${\cal N}=1$ vector multiplet
(\ref{n1vector}), and define $\sI_{X}^{\rm tensor}$ by
\begin{align}
\sI^{\rm tensor}_{X}=\frac{uq}{1-uq}.
\label{itensor}
\end{align}
In \cite{Nakayama:2005mf} this was extracted from
the index of the gauge theory associated with the orbifold.
We can generalize this to other fixed loci appearing along intersections of $\bm{S}^5$ and $X_I$ planes.
We denote the corresponding single-particle
index by $\sI_I^{\rm tensor}$,
and it is given by
\begin{align}
\sI^{\rm tensor}_{I}
=\frac{u_Iq}{1-u_Iq}.
\end{align}
Interestingly, these are the same as terms appearing in $\sI^{\rm grav}$ in (\ref{igravdea}).

If $\ZZ_n$ is a proper subgroup of $\wt\Gamma$
and $\wt\Gamma$ includes elements that act non-trivially on
the fixed locus, we need to perform the corresponding projection.
It is realized by the projection ${\cal P}$ defined in (\ref{pgammadef}).
In a general situation in which
the orbifold has
type $A_{n_I-1}$ singular locus on the $X_I$-plane
the total contribution of the
tensor multiplets is given by
\begin{align}
{\cal I}^{\rm tensor}
=
\prod_{I=X,Y,Z}({\cal I}^{\rm tensor}_{I})^{n_I-1},\quad
{\cal I}^{\rm tensor}_{I}&=\Pexp\left({\cal P}\sI^{\rm tensor}_{I}\right).
\end{align}

%%%%%%%%%%%%%%%%%%%%%%%%%%%%%%%%%%%%
\section{Finite $N$ corrections from wrapped D3-branes}\label{finiten.sec}
The purpose of this section
is to give the prescription to calculate
the leading finite $N$ corrections
to the index from wrapped D3-branes based on some assumptions.
The wrapping number of D3-branes
in ${\cal Y}$ is related to the baryonic charges.
We first define
the wrapping number and then
establish the relation to the
baryonic charges.

%%%%%%%%%%%%%%%%%%%%%%%%%
\subsection{Wrapping number}
Let us define $\bm{S}^5$ as the subset of $\CC^3$ by
\begin{align}
|X|^2+|Y|^2+|Z|^2=1
\end{align}
and represent the worldvolume of a D3-brane in $\bm{S}^5$ as
the intersection of a surface
\begin{align}
F(X,Y,Z)=0
\label{fzero}
\end{align}
in $\CC^3$ and the $\bm{S}^5$.
For the D3-brane to be BPS the function $F$ must be holomorphic \cite{Mikhailov:2000ya}.
Here we are interested in topological aspects and
the holomorphy is not assumed.

When we consider D3-branes in the orbifold ${\cal Y}=\bm{S}^5/\wt\Gamma$
the surface (\ref{fzero}) must be invariant under the orbifold action.
This requires the invariance of function $F$ up to
an overall factor:
\begin{align}
F(w_X(\wt g)X,w_Y(\wt g)Y,w_Z(\wt g)Z)=w(\wt g)F(X,Y,Z)\quad
\forall\wt g\in\wt\Gamma.
\label{finv}
\end{align}
where $w_I(\wt g)$ are diagonal elements in the matrix form (\ref{c3action}) of $\wt\Gamma$
and $w(\wt g)$ is a phase factor depending on $\wt g$.
Note that $w$ as well as $w_I$ is an element of the dual group (\ref{dualgroup}).
A continuous deformation of the worldvolume is realized by changing the function $F$ continuously
without violating the relation (\ref{finv}).
Because $w$ takes discrete values, it does not change under continuous deformations.
Namely, $w$ is a topological invariant associated with the D3-brane worldvolume
in ${\cal Y}$.
We call $w$ the wrapping number (although it is not an integer but an element of $\Gamma$).

We will later focus on brane configurations
described by monomial holomorphic functions $F(X_I)$.
If $F(X_I)$ is such a function,
by setting $X=Y=Z=1$ in (\ref{finv}) we obtain
\begin{align}
w=F(w_X,w_Y,w_Z).
\end{align}
This relation directly gives the
wrapping number of the configuration given by
the function $F$.
In particular,
we obtain $w=w_I$ for $F=X_I$.
Namely, the diagonal components
in the orbifold action (\ref{c3action})
are nothing but the wrapping numbers
of three brane configurations
$X_I=0$ which will play
a central role in the following analysis.

If the orbifold has fixed loci there are shrinking two-cycles along
the loci.
The above definition of the wrapping number
does not take these shrinking cycles into account.
D3-branes wrapped around these cycles
are regarded as tensionless strings in the $(2,0)$ theories.
The analysis in \cite{Nakayama:2005mf} showed that the index of the gauge theory
in the large $N$ limit is reproduced as the contribution
of the gravity multiplet in the bulk and the tensor multiplets
localized on the loci,
and the tensionless strings do not contribute the index.

To fully specify a D3-brane classical configuration
we need to give not only the shape of the worldvolume but also the gauge field
on the brane.
Let us consider a D3-brane wrapped on $X_I=0$.
The topology of the worldvolume is ${\cal C}=\bm{S}^3/\wt\Gamma$,
which may have fixed loci.
The flat connection on ${\cal C}_*$
is classified by
$H^1({\cal C},U(1))=\Gamma$.%
\footnote{If ${\cal C}$ contains an $A_{n+1}$ type singular locus
it should be treated in this equation as a curve with mod $n$ linking.
This is because if we move an endpoint of an open string attached on the worldvolume around
the singular locus $n$ times the resulting open string can be continuously deformed back into the original configuration.}

For general monomial functions
\begin{align}
F(X_I)=\prod_{I=X,Y,Z}(X_I)^{n_I},
\end{align}
the equation $F(X_I)=0$ gives a set of three stacks of
D3-branes.
A stack wrapped on $X_I=0$ contains $n_I$ D3-branes,
and $U(n_I)$ gauge theory is realized on it.
We can specify a flat $U(1)^{n_I}\subset U(n_I)$ connection
on each stack by a set of $n_I$ elements of $\Gamma$:
\begin{align}
h_I=\{h_{I,1},h_{I,2},\ldots,h_{I,n_I}\},\quad
h_{I,i}\in\Gamma.
\end{align}
These holonomy variables are shifted by the global $1$-form
symmetry associated with the background NS-NS two-form field,
which we call $T$-symmetry.
This symmetry is parameterized by $t\in H^1({\cal Y}_*,U(1))=\Gamma,$\footnote{${\cal Y}_*$ is defined from ${\cal Y}$ by removing fixed loci.}
and it acts on the holonomy variables
as $h_{I,i}\rightarrow th_{I,i}$.

%%%%%%%%%%%%%%%%%%%%%%%%%%%%%%%%%%%%%%%%%%%%%%%%%%%
\subsection{Wrapped D3-branes and baryonic charges}
Based on the results of the analysis of S-fold theories in \cite{Arai:2019xmp}
we assume that the finite $N$ corrections can be obtained from
D3-branes
wrapped around particular three-cycles
$X=0$, $Y=0$, and $Z=0$.

The relation between wrapped D3-branes and baryonic operators was first pointed out in \cite{Witten:1998xy}.
In general a wrapped D3-brane has non-trivial topology,
and then we need to take account of the holonomy of the
gauge field on the brane \cite{Gukov:1998kn}.
We can easily generalize the relation between wrapped branes and baryonic operators
to a general orbifold.

Let us consider a D3-brane wrapped over one of the cycles $X_I=0$.
The flat connection on the worldvolume is specified by one holonomy variable
$h\in\Gamma$.
Namely, the classical configuration is specified by $I$ and $h$.
On the gauge theory side we have the corresponding baryonic operator\footnote
{Precisely, there is an ambiguity associated with
the choice of origin of the holonomy variables,
and (\ref{bihdef}) should be replaced with
${\cal B}_{I,h}=\det(\Phi_I^{c_Ih\rightarrow c_Iw_Ih})$,
where $c_I$ are elements of $\Gamma$ depending on the choice of the origin.
We cannot fix $c_I$ only by the consistency with the $T$-symmetry.
For each $I$ the choice of $c_I$ is a matter of convention.
However, relative values amomg $c_I$ are physical.
In this paper we focus on configurations
with a single wrapped brane and the corresponding indices are not affected by $c_I$.
For this reason we simply neglect this ambiguity in the following.
}
\begin{align}
{\cal B}_{I,h}=\det(\Phi_I^{h\rightarrow w_Ih}).
\label{bihdef}
\end{align}
This identification is consistent with the fact that
on the gauge theory side we can identify the $T$-symmetry as the shift symmetry of the
periodic quiver diagram.
(We assume that the marginal deformation parameters such as gauge couplings are
appropriately tuned.)
An element $t\in \Gamma$ acts on the basis vectors of the
baryonic charges as
$\bm{b}_g\rightarrow\bm{b}_{tg}$.

As a simple check we can easily confirm that the dimension
$\dim{\cal B}_{I,h}=N$ is reproduced as the mass of the D3-brane.
For the orbifold ${\cal Y}=\bm{S}^5/\wt\Gamma$
the common radius $L$ of the $\bm{S}^5$ and the $AdS_5$
is given in terms of the D3-brane tension $T_{\rm D3}$ by
\begin{align}
L^4=\frac{|\Gamma|N}{2\pi^2T_{\rm D3}}.
\end{align}
The mass of a D3-brane wrapped over one of the cycles $X_I=0$
is $M=2\pi^2L^3T_{\rm D3}/|\Gamma|=N/L$,
and the dimension of the corresponding operator is $LM=N$.
We can also show that the wrapped brane carries the correct R-charges
$(R_I,R_{I+1},R_{I+2})=(N,0,0)$
\footnote{We treat $I=X,Y,Z$ as a cyclic variable. For example, if $I=Z$, $(R_I,R_{I+1},R_{I+2})=(R_Z,R_X,R_Y)$.}
 due to the coupling to the
background R-R flux \cite{McGreevy:2000cw}.

We can generalize this correspondence to the configuration
described by a general monomial function
$F(X,Y,Z)=X^{n_X}Y^{n_Y}Z^{n_Z}$.
On the gravity side $F(X,Y,Z)=0$ gives
$n_X$ D3-branes wrapped on $X=0$,
$n_Y$ D3-branes wrapped on $Y=0$,
and $n_Z$ D3-branes wrapped on $Z=0$.
Let $\vec h_I=\{h_{I,1},\ldots,h_{I,n}\}$
be the set of $n_I$ elements of $\Gamma$
representing the holonomy on the branes wrapped on $X_I=0$.
A corresponding operator is
\begin{align}
{\cal O}
=\prod_{I=X,Y,Z}\prod_{i=1}^{n_I}{\cal B}_{I,h_{I,i}}.
\label{operators}
\end{align}
By reading off the baryonic charges
from (\ref{operators}) we obtain
the map from the
brane configuration with the holonomy on it
to the baryonic charges:
\begin{align}
\bm{b}
=\sum_{I=X,Y,Z}\sum_{i=1}^{n_I}
(\bm{b}_{h_{I,i}}-\bm{b}_{w_Ih_{I,i}}).
\end{align}

Now let us establish the relation between the wrapping number $w\in\Gamma$
and the baryonic charge $\bm{b}\in\Lambda_B$.
In general $\Lambda_B$ is larger than $\Gamma$ and we cannot simply identify
$\bm{b}$ and $w$.
This difference comes from
the holonomy degrees of freedom
on wrapped D3-branes.
To relate the baryonic charges to
the wrapping number
we need to eliminate the information associated with the $T$-symmetry
from the baryonic charges.
Let us define the reduced lattice $\Lambda_B^{\rm red}$ from $\Lambda_B$ by
forgetting about the information associated with $T$ symmetry:
\begin{align}
\Lambda_B^{\rm red}=\Lambda_B/\stackrel{T}{\sim},
\label{wandbsim}
\end{align}
where
the equivalence relation $\stackrel{T}{\sim}$ is defined by
\begin{align}
\bm{b}_{g_1}-\bm{b}_{g_2}
\stackrel{T}{\sim}\bm{b}_{tg_1}-\bm{b}_{tg_2}\quad
t,g_1,g_2\in\Gamma.
\label{tidentification}
\end{align}
In fact,
we can easily show $\Delta\bm{b}_g\stackrel{T}{\sim}0$
and hence
\begin{align}
\Lambda^{\rm red}_B
=\Lambda_B/\stackrel{T}{\sim}\
=\Lambda_B^0/\stackrel{T}{\sim}.
\end{align}
Up to the $T$-symmetry
the sum of two basis vectors of the classically conserved baryonic charge
lattice $\Lambda_B^0$
is given by
\begin{align}
(\bm{b}_e-\bm{b}_{g_1})
+(\bm{b}_e-\bm{b}_{g_2})
&\stackrel{T}{\sim}
(\bm{b}_e-\bm{b}_{g_1})
+(\bm{b}_{g_1}-\bm{b}_{g_1g_2})
=
\bm{b}_e-\bm{b}_{g_1g_2}.
\end{align}
This relation implies that
the map $\bm{b}_e-\bm{b}_g\rightarrow g$
is an isomorphism,
and $\Lambda_B^{\rm red}\cong\Gamma$.
We can simply identify
$\Lambda^{\rm red}_B$ with the group of the wrapping number.
The homomorphism $\varphi$ from
$\Lambda_B^0$ to $\Lambda_B^{\rm red}$ is given by
\begin{align}
\varphi:\bm{b}=\sum_{g\in\Gamma}B_g\bm{b}_g\rightarrow w=\prod_{g\in\Gamma}g^{-B_g}.
\end{align}

The index for a specific $w$ is given by
\begin{align}
{\cal I}_w=\sum_{\bm{b}}{\cal I}_{\bm{b}},
\label{iwgauge}
\end{align}
where the summation is taken over $\bm{b}$ satisfying $\varphi(\bm{b})=w$.
Just like ${\cal I}_{\bm b}$ and ${\cal I}_{\wh{\bm{b}}}$
we can extract ${\cal I}_w$ from the total index ${\cal I}$
by the $\zeta$-expansion.
For this purpose we impose the constraint
\begin{align}
\frac{\zeta_{g_1}}{\zeta_{g_2}}=
\frac{\zeta_{tg_1}}{\zeta_{tg_2}},\quad
\forall g_1,g_2,t\in\Gamma
\label{crossratio}
\end{align}
corresponding to (\ref{tidentification}).
Then $\zeta^{\bm{b}}$ depends on $\bm{b}$ through the
corresponding wrapping number $w=\varphi(\bm{b})$
and we denote it by $\zeta^w$.
We can rewrite (\ref{clexpansion}) as
\begin{align}
{\cal I}=\sum_{w\in\Gamma}\zeta^w{\cal I}_w.
\label{wexpansion}
\end{align}
Because of (\ref{total0}) ${\cal I}$ is invariant under the overall
phase rotation $\zeta_g\rightarrow e^{i\theta}\zeta_g$.
With this redundancy we can set $\zeta_e=1$.
Then the relation (\ref{crossratio}) means that
the map $g\rightarrow\zeta_g$ is a homomorphism.
Namely, we can regard the fugacities
satisfying (\ref{crossratio}) as an element of the
orbifold group $\wt\Gamma$.
$\zeta^w$ in (\ref{wexpansion}) is nothing but the pairing
of $\zeta\in\wt\Gamma$ and $w\in\Gamma$.

%%%%%%%%%%%%%%%%%%%%%%%%%
\subsection{Index from wrapped D3-branes}

In the next section we calculate indices on the both sides of the duality
for each sector specified by the wrapping number $w\in\Gamma$ and compare them.
For distinction,
we denote the index on the gauge theory side by ${\cal I}^{\rm gauge}_w$ and the
index on the gravity side by ${\cal I}^{\rm AdS}_w$.

On the gravity side
the index
${\cal I}_w^{\rm AdS}$ is given by
\begin{align}
{\cal I}_w^{\rm AdS}={\cal I}^{\rm KK}\sum_F\sum_h{\cal I}^{\rm D3}_{F,h},
\end{align}
where
${\cal I}^{\rm D3}_{F,h}$ is the index of D3-branes with
the worldvolume $F=0$ and the holonomy $h$ on it.
The summation is taken over monomials $F$ satisfying $F(w_I)=w$
and holonomies $h$ for each wrapped D3-brane configuration $F=0$.
If $F$ is a monomial of order $k$ the index is ${\cal O}(q^{kN})$.
If $k=1$ the holonomy varianble is a single element of $\Gamma$,
while if $k\geq2$ $h$ is a set of $k$ elements of $\Gamma$.

Because we are interested in the leading finite $N$ corrections we focus on
the four monomials $F=1,X,Y$, and $Z$,
which give contribution to the sectors with $w=e$, $w_X$, $w_Y$, and $w_Z$,
respectively.
For $w\neq e, w_I$ the index ${\cal I}_w$
is of order ${\cal O}(q^{2N})$ or higher,
and we will not pay attention to them.

Let us first consider the case that
$e$ and $w_I$ are all different.

The leading contribution to ${\cal I}^{\rm AdS}_e$ is given by $F=1$, which
gives no wrapped D3-branes and the index includes only
the Kaluza-Klein contributions
\begin{align}
{\cal I}_e^{\rm AdS}={\cal I}^{\rm KK}+{\cal O}(q^{kN}),
\label{inde}
\end{align}
where $k$ is the lowest order of
non-trivial monomial satisfying $F(w_I)=e$.

On the gauge theory side
the baryonic charge $\bm{b}=0$ gives the leading contribution to ${\cal I}_e^{\rm gauge}$;
\begin{align}
{\cal I}^{\rm gauge}_e={\cal I}^{\rm gauge}_0+{\cal O}(q^{kN}),
\label{indegg}
\end{align}
where $k$ is the same as in (\ref{inde}).
By comparing
(\ref{inde}) and (\ref{indegg})
we obtain
\begin{align}
{\cal I}_0^{\rm gauge}={\cal I}^{\rm KK}+{\cal O}(q^{kN}).
\label{inde2}
\end{align}

The leading contribution to ${\cal I}^{\rm AdS}_{w_X}$ is given by
$F=X$, which gives a single D3-brane wrapped over $X=0$.
The index on the gravity side is given by
\begin{align}
{\cal I}^{\rm AdS}_{w_X}
&={\cal I}^{\rm KK}\sum_h{\cal I}^{\rm D3}_{X,h}+{\cal O}(q^{kN}),
\end{align}
where $k$ is the lowest order of monomials $F\neq X$
satisfying $F(w_I)=w_X$.
The worldvolume theory on the single D3-brane is a $U(1)$ gauge theory
and no fields couple to the holonomy.
Therefore,
the index ${\cal I}_{X,h}$ is independent of $h$
and we can replace the summation over $h$ by the factor $|\Gamma|$:
\begin{align}
{\cal I}_{w_X}^{\rm AdS}
&=|\Gamma|{\cal I}^{\rm KK}{\cal I}^{\rm D3}_{X,0}+{\cal O}(q^{kN}).
\label{indxys}
\end{align}

On the gauge theory side the baryonic charges of sectors giving leading contribution
are $\bm{b}_g-\bm{b}_{w_Xg}$.
Due to the $T$-symmetry
the index ${\cal I}^{\rm gauge}_{\bm{b}_g-\bm{b}_{w_Xg}}$ does not depend on $g$,
and this degeneracy gives the factor $|\Gamma|$:
\begin{align}
{\cal I}_w^{\rm gauge}
=|\Gamma|{\cal I}^{\rm gauge}_{\bm{b}_e-\bm{b}_{w_X}}+{\cal O}(q^{kN}).
\label{indxgg}
\end{align}
By comparing (\ref{indxys}) and (\ref{indxgg})
we obtain
\begin{align}
|\Gamma|{\cal I}_{\bm{b}_e-\bm{b}_{w_X}}
=|\Gamma|{\cal I}^{\rm KK}{\cal I}^{\rm D3}_{X,0}+{\cal O}(q^{kN}).
\label{indxys2}
\end{align}
We also have similar relations for $w=w_Y$ and $w=w_Z$.

The index for a D3-brane wrapped on $X_I=0$ was calculated in \cite{Arai:2019xmp}
for S-fold theories, and the result can be used for orbifolds, too, by simply
replace the S-fold projection operator used in
\cite{Arai:2019xmp} by the orbifold projection defined in (\ref{pgammadef}).
${\cal I}_{X_I,0}^{\rm D3}$ are given by
\begin{align}
{\cal I}_{X_I,0}^{\rm D3}
=
q^Nu_I^N
\Pexp({\cal P}\sI_I^{\rm D3}),
\end{align}
where
the single-particle indices are the same as what are used in \cite{Arai:2019xmp}:
\begin{align}
\sI^{{\rm D3}}_{I}&=
\frac{
\frac{1}{u_Iq}
-(y+\frac{1}{y})\frac{1}{u_I}q^{\frac{1}{2}}
-(u_{I+1}+u_{I+2})q
+(y+\frac{1}{y})q^{\frac{3}{2}}
+2\frac{1}{u_I}q^2
-q^3
}{(1-u_{I+1}q)(1-u_{I+2}q)}.
\label{threeid3}
\end{align}
(We set the $U(1)_A$ fugacity $\eta$ introduced in \cite{Arai:2019xmp} to realize the S-fold action to be $1$.)

The relations in (\ref{inde2}) and (\ref{indxys2}) hold
only when $e,w_X,w_Y,w_Z$ are all different.
If some of them coincide,
the corresponding index ${\cal I}_w$ is given as the sum of
their contributions.

Let us consider the case that one of $w_I$, say $w_X$, is $e$.
This is the case
for the orbifold $\CC\times\CC^2/\ZZ_n$ with $\ZZ_n$ acting on $Y$ and $Z$.
In this case the two configurations given by $F=1$ and $F=X$ contribute to ${\cal I}_e$:
\begin{align}
{\cal I}_e
={\cal I}^{\rm KK}
(1+|\Gamma|{\cal I}^{\rm D3}_{X,0}+\cdots).
\end{align}
$w_X=e$ means that the wrapping number of the configuration
$X=0$ is trivial and
we can unwrap the brane by a continuous deformation.
Correspondingly, the single particle index includes
a term proportional to $q^{-1}$,
which we call the tachyonic term.
\begin{align}
{\cal P}\sI_X^{\rm D3}=\frac{1}{uq}+\cdots.
\end{align}
When we calculate the plethystic exponential such a tachyonic term
is treated as follows.
\begin{align}
\Pexp\left(\frac{1}{uq}\right)
=\frac{1}{1-\frac{1}{uq}}
=-\frac{uq}{1-uq}
=-uq\Pexp(uq).
\end{align}
Although the physical meaning of this is not clear
it was found in \cite{Arai:2019xmp} that this gives the correct leading
finite $N$ corrections for ${\cal N}=4$ SYM,
and we simply assume that this works for orbifolds, too.

If $w=w_Y=w_Z\neq e$, the system has $SU(2)$ flavor symmetry
mixing $Y$ and $Z$.
This is the case if all dots in the toric diagram except ones at corners are
aligned on a median.
The two configurations $Y=0$ and $Z=0$
contribute to the $w$ sector;
\begin{align}
{\cal I}_w={\cal I}^{\rm KK}
(|\Gamma|{\cal I}^{\rm D3}_{Y,0}+|\Gamma|{\cal I}^{\rm D3}_{Z,0}+\cdots)
\end{align}
In this case, two configurations $Y=0$ and $Z=0$ belong to the
same homology class,
and we can continuously deform them to each other via the intermediate
configurations given by
\begin{align}
bY+cZ=0.
\label{cp1config}
\end{align}
The coefficients $b$ and $c$ are the homogeneous coordinates
of the configuration space $\CC\bm{P}^1$.
In this case each of the single particle indices
${\cal P}\sI_Y^{\rm D3}$ and
${\cal P}\sI_Z^{\rm D3}$ includes a term proportional to $q^0$.
We call such terms ``zero-mode terms.''
\begin{align}
{\cal P}\sI_Y^{\rm D3}=\frac{u}{v^2}+\cdots,\quad
{\cal P}\sI_Z^{\rm D3}=\frac{v^2}{u}+\cdots.
\end{align}
The plethystic exponentials of these zero-mode terms
give the fractional factors
\begin{align}
{\cal I}_{Y,0}^{\rm D3}=q^N\frac{v^N}{u^N}\Pexp{\cal P}\sI_Y^{\rm D3}&=\frac{(v/u)^N}{1-u/v^2}\left(q^N+\cdots\right),\nonumber\\
{\cal I}_{Z,0}^{\rm D3}=q^N\frac{1}{v^N}\Pexp{\cal P}\sI_Z^{\rm D3}&=\frac{1/v^N}{1-v^2/u}\left(q^N+\cdots\right).
\end{align}
These are combined into the character of the $SU(2)$ flavor symmetry:
\begin{align}
{\cal I}_{Y,0}^{\rm D3}+{\cal I}_{Z,0}^{\rm D3}
&=\left(\frac{v^N}{u^N}+\frac{v^{N-2}}{u^{N-1}}+\cdots+\frac{1}{v^N}\right)q^N+\cdots
\nonumber\\
&=\chi_N(\tfrac{v}{u},\tfrac{1}{v})q^N+\cdots,
\end{align}
where $\chi_N$ is the $u(2)$ character defined in (\ref{u2charactor}).

If $w=w_X=w_Y=w_Z$ are satisfied and
the system has the flavor symmetry $SU(3)$.
This is the case for $\CC^3$
and $\CC^3/\ZZ_3$.
For $\CC^3/\ZZ_3$ the index of the $w$-sector
is given as the sum of three contributions
up to higher order corrections:
\begin{align}
{\cal I}_w
={\cal I}^{\rm KK}
(|\Gamma|{\cal I}^{\rm D3}_{X,0}+|\Gamma|{\cal I}^{\rm D3}_{Y,0}+|\Gamma|{\cal I}^{\rm D3}_{Z,0}+\cdots).
\label{ixyzsub}
\end{align}
(For $\CC^3$, $w_X=w_Y=w_Z=e$ and we have additional $1$'' in the parentheses on the right hand side
in (\ref{ixyzsub}).)
In this case
the three wrapped brane configurations are continuously deformed among them
through the intermediate configurations
\begin{align}
aX+bY+cZ=0.
\end{align}
$a$, $b$, and $c$ are homogeneous coordinates in the
configuration space $\CC\bm{P}^2$.
Corresponding to the continuous deformation
each of $\sI_{X_I}^{\rm D3}$
has two zero-mode terms.
The corresponding fractional factors give the $SU(3)$ character
when the three contributions are summed.

%%%%%%%%%%%%%%%%%%%%%%%%%%%%%%%%%%%%%%%
\section{Examples}\label{examples.sec}
In this section we confirm that by the prescription given in the
previous section we can obtain the leading finite $N$ corrections
for $N=2$.
The results for $N=3$ are shown in Appendix \ref{n3.sec}.

%%%%%%%%%%%%%%%%%%%%%%%%%%%%%%%%%%%%%%%%%%%%%%%%%%%%%%%%%%%%%%%
\subsection{$\mathbb{C}^3/\mathbb{Z}_{2n+1}$ ($n=1,2,3,\ldots$)}
Let us first consider the series of orbifolds ${\cal X}=\CC^3/\ZZ_{2n+1}$ ($n=1,2,3,\ldots$)
with $\ZZ_{2n+1}$ generated by $\diag(\omega_{2n+1}^{-2},\omega_{2n+1},\omega_{2n+1})$.
The dual group $\Gamma=\ZZ_{2n+1}$ is characterized by
the relations
\begin{align}
w_Y=w_Z,\quad
w_X=w_Y^{-2},\quad
w_Y^{2n+1}=e.
\end{align}
The $n=1$ case is special in the sense that $w_X$ agrees with $w_Y=w_Z$ then and the flavor
symmetry is enhanced to $SU(3)$.
We give the results for $n=1,2$, and $3$.

\subsubsection{$\mathbb{C}^3/\mathbb{Z}_3$}
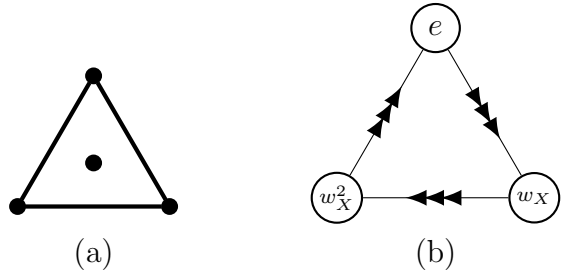
\begin{figure}[htbp]
\begin{center}
\begin{center}
\begin{tikzpicture}
%Toric diagram
   %Nodes
    \node [draw=black, circle, scale=0.5, thick, fill=black] (toric1) at (-3.5,{-3/2*1/sqrt(3)}) {};
    \node [draw=black, circle, scale=0.5, thick, fill=black] (toric2) at ($(toric1)+(-1,{sqrt(3)})$) {};
    \node [draw=black, circle, scale=0.5, thick, fill=black] (toric3) at ($(toric1)+(-2,0)$) {};
    \node [draw=black, circle, scale=0.5, thick, fill=black] (toric4) at ($(toric1)+(-1,{1/sqrt(3)})$) {};
    %Arrows
    \draw [ultra thick](toric1) -- (toric2);
    \draw [ultra thick](toric2) -- (toric3);
    \draw [ultra thick](toric3) -- (toric1);

%Quiver diagram
    %Nodes
    \node [draw=black, circle, inner sep=3.8pt, thick, fill=white] (SU1) at (90:1.5) {$e$};
    \node [draw=black, circle, inner sep=1.8pt, thick, fill=white] (SU2) at (90-360/3:1.5) {\scriptsize{$w_X$}};
    \node [draw=black, circle, inner sep=1pt, thick, fill=white] (SU3) at (90-2*360/3:1.5) {\scriptsize{$w_X^2$}};
    %Arrows
    \draw [->>>-] (SU1) -- (SU2);
    \draw [->>>-] (SU2) -- (SU3);
    \draw [->>>-] (SU3) -- (SU1);

\node [] at (-4.5,-1.5) {(a)};
\node [] at (0,-1.5) {(b)};
\end{tikzpicture}
\end{center}
\caption{The toric diagram of $\CC^3/\ZZ_3$ (a) and the corresponding quiver diagram (b) are shown.}
\label{quiverz3}
\end{center}
\end{figure}

If $n=1$ the relations among $w_I$ are
\begin{align}
w_X=w_Y=w_Z,\quad
w_X^3=e.
\label{wrelz3}
\end{align}
The toric diagram and the quiver diagram are shown in figure \ref{quiverz3}.
There are three sectors labeled by
$w=e, w_X, w_X^2$. (See Table \ref{z3.tbl}.)
\begin{table}[htb]
\caption{Wrapping numbers $w\in\Gamma$ and the corresponding monomials $F(X_I)$ are shown
for the orbifold ${\cal X}=\CC^3/\ZZ_3$. The shaded row is irrelevant to the leading corrections.}
\label{z3.tbl}
\vspace{3mm}
\centering
\begin{tabular}{ccc}
\hline
\hline
$w$ & ${\cal O}(q^0)$ and ${\cal O}(q^N)$ & higher order \\
\hline
$e$         & $1$           & $X^3$, ... \\
$w_X$       & $X$, $Y$, $Z$ & $X^4$, ... \\
\sh $w_X^2$ & \sh           &\sh $X^2$, ... \\
\hline
\end{tabular}
\end{table}

Let us first consider the sector with trivial wrapping $w=e$, which we call the mesonic sector.
The relations (\ref{wrelz3}) show that
any monomial $w_X^{n_1}w_Y^{n_2}w_Z^{n_3}$ of order $n_1+n_2+n_3=3k$ ($k=0,1,2,\ldots$) is $e$,
and the corresponding function $F(X,Y,Z)=X^{n_1}Y^{n_2}Z^{n_3}$ gives a brane configuration
$F=0$ contributing to the index of the mesonic sector.
Such a monomial with the lowest order is $F=1$.
The corresponding D3-brane configuration, $F=0$, does not contain any D3-brane,
and the corresponding index is ${\cal I}^{\rm D3}_{1}=1$.
We show this monomial in the second column in Table \ref{z3.tbl}.
There exist infinite number of higher order monomials which give
higher order contributions to the index.
An example of such higher order monomials with the lowest order, $X^3$, is shown in
the third column in Table \ref{z3.tbl}.
This higher order contribution is ${\cal O}(q^{3N})$, and hence we have the relation
${\cal I}_e^{\rm D3}=1+{\cal O}(q^{3N})$.
On the gauge theory side, the operator corresponding to $F=1$ is the identity operator
with the baryonic charge $\bm{b}=0$.
For the higher order monomial like $F=X^3$, there are many operators carrying different baryonic charges
depending on the holonomy.
Because such higher order operators give indices of order ${\cal O}(q^{3N})$ or higher,
we obtain the relation ${\cal I}_e^{\rm gauge}={\cal I}_0^{\rm gauge}+{\cal O}(q^{3N})$.
Combining these relations we obtain the relation
\begin{align}
{\cal I}_0^{\rm gauge}&={\cal I}^{\rm grav}+{\cal O}(q^{3N}),
\label{z3mesonic}
\end{align}
which we want to check.
We calculate the ${\cal I}_0^{\rm gauge}$ on the left hand side for small $N$ by the localization formula.
The result for $N=2$ is
\begin{align}
\mathcal{I}^{\rm gauge}_0
&=1+\left(1-\chi_{(1,1)}+\chi_{(3,0)}\right)q^3
\nonumber\\
&\quad +\left(-2+\chi_{(0,3)}-\chi_{(1,1)}-\chi_{(3,0)}-2\chi_{(4,1)}+2\chi_{(6,0)}\right)q^6
+\mathcal{O}(q^{\frac{15}{2}}).
\label{z3mgauge}
\end{align}
${\cal I}^{\rm grav}$ on the right hand side
in (\ref{z3mesonic})
is obtained by the formula (\ref{igravdef}).
It is independent of $N$ and given by
\begin{align}
\mathcal{I}^{\mathrm{grav}}
&=1+\left(1-\chi_{(1,1)}+\chi_{(3,0)}\right)q^3
\nonumber\\
&\quad +\left(1+\chi_{(0,3)}-\chi_{(1,1)}+2\chi_{(3,0)}-2\chi_{(4,1)}+2\chi_{(6,0)}\right)q^6
+\mathcal{O}(q^{9})
\label{z3mgrav}
\end{align}
The comparison of (\ref{z3mgauge}) and (\ref{z3mgrav}) shows that
the relation (\ref{z3mesonic}) certainly holds for $N=2$.

Next, let us consider the sectors with $w\neq e$.
We call such sectors baryonic sectors.
In the example of $\CC^3/\ZZ_3$ we have two baryonic sectors
with $w=w_X$ and $w=w_X^2$.
For the former
the corresponding functions $F$ with the lowest order are $F=X$, $F=Y$, and $F=Z$,
which are shown in the second column in Table \ref{z3.tbl}.
For each of D3-brane configurations $X=0$, $Y=0$, and $Z=0$,
the holonomy on the brane is specified by an element of $\Gamma=\ZZ_3$.
We should sum up contributions of three D3-brane configurations
with all possible holonomies.
They gives the index of order ${\cal O}(q^N)$.
We also have higher order functions $F$.
Among such functions an example with the lowest order, $X^4$, is shown in the third column
in Table \ref{z3.tbl}.
The D3-branes corresponding to such higher order functions give ${\cal O}(q^{4N})$ contributions to the
index.
Therefore, the D3-brane contribution for the $e=w_X$ sector is
${\cal I}^{\rm D3}_{w_X}=3{\cal I}^{\rm D3}_{X,0}+3{\cal I}^{\rm D3}_{Y,0}+3{\cal I}^{\rm D3}_{Z,0}+{\cal O}(q^{4N})$.
On the gauge theory side, the operators corresponding to the brane configuration $X_I=0$ are
$\det(\Phi_I^{g\rightarrow w_Xg})$ ($g\in\ZZ_3$), which have
the baryonic charges $\bm{b}=\bm{b}_g-\bm{b}_{w_Xg}$.
Three sectors labeled by these baryonic charges
give the same index due to the $T$-symmetry,
and we obtain the relation
${\cal I}^{\rm gauge}_{w_X}=3{\cal I}_{\bm{b}_e-\bm{b}_{w_X}}+{\cal O}(q^{4N})$.
The discrepancy of order ${\cal O}(q^{4N})$ comes from the existence
of other baryonic charges corresponding to $w=w_X$.
By combining the relations above we obtain
\begin{align}
\frac{{\cal I}^{\rm gauge}_{\bm{b}_e-\bm{b}_{w_X}}}{{\cal I}^{\rm gauge}_0}
&={\cal I}^{\rm D3}_{X,0}+{\cal I}^{\rm D3}_{Y,0}+{\cal I}^{\rm D3}_{Z,0}+{\cal O}(q^{4N}).
\label{b1z3}
\end{align}

Let us confirm this relation by the numerical calculation for $N=2$.
The result of the localization for the left hand side of
(\ref{b1z3}) is
\begin{align}
\frac{{\cal I}^{\rm gauge}_{\bm{b}_e-\bm{b}_{w_X}}}{{\cal I}^{\rm gauge}_0}
&=\chi_{(2,0)}q^2
+\chi^J_1\left(\chi_{(2,0)}-\chi_{(0,1)}\right)q^{\frac{7}{2}}
\nonumber\\
&\quad +\left(\chi^{J}_2\left(\chi_{(2,0)}-\chi_{(0,1)}\right)-\left(\chi_{(0,1)}+\chi_{(1,2)}+\chi_{(2,0)}\right)\right)q^5\nonumber\\
&\quad +\left(\chi^J_3\left(\chi_{(2,0)}-\chi_{(0,1)}\right)+\chi^J_1\left(\chi_{(3,1)}-\chi_{(1,2)}\right)\right)q^{\frac{13}{2}}\nonumber\\
&\quad +\left(4\chi_{(1,2)}+3\chi_{(2,0)}+\chi^J_4\chi_{(2,0)}+\chi_{(2,3)}+3\chi_{(3,1)}+\chi_{(5,0)}-\chi_{(6,1)}\right.
\nonumber\\
&\hspace{2cm}
\left.+\left(7+\chi^J_2-\chi^J_4\right)\chi_{(0,1)}\right)q^8
+\mathcal{O}(q^{\frac{19}{2}}),
\label{Z3N2gauge}
\end{align}
The right hand side of (\ref{b1z3}) is
\begin{align}
&{\cal I}^{\rm D3}_{X,0}+{\cal I}^{\rm D3}_{Y,0}+{\cal I}^{\rm D3}_{Z,0}
=\cdots(\mbox{identical terms})\cdots
\nonumber\\
&\quad +\left(\chi_{(1,2)}+\chi_{(2,0)}+\chi^J_4\chi_{(2,0)}+\chi_{(2,3)}+\chi_{(3,1)}-\chi_{(5,0)}-\chi_{(6,1)}\right.\nonumber\\
&\hspace{2cm} \left.+\left(4+\chi^J_2-\chi^J_4\right)\chi_{(0,1)}\right)q^8
+\mathcal{O}(q^{\frac{19}{2}}).
\label{Z3N2gr}
\end{align}
In this equation ``$\cdots(\mbox{identical terms})\cdots$" mean that
the terms are the same as the corresponding terms in (\ref{Z3N2gauge}).
Comparison of
(\ref{Z3N2gauge}) and
(\ref{Z3N2gr}) confirm the relation (\ref{b1z3}).

For the other baryonic sector with the wrapping number $e=w_X^2$
the index is ${\cal O}(q^{2N})$.
We do not analyze such sectors in this paper.

%%%%%%%%%%%%%%%%%%%%%%%%%%%%%%%%%%%%%%%%%%%%%%%%%%%%%%%%%%%%%%%%%%%%%%%%%%%
\subsubsection{$\CC^3/\ZZ_5$}
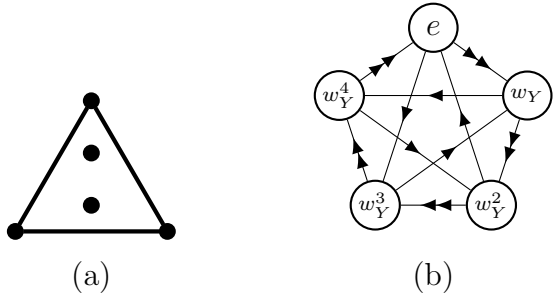
\begin{figure}[htbp]
\begin{center}
\begin{tikzpicture}
%Toric diagram
   %Nodes
    \node [draw=black, circle, scale=0.5, thick, fill=black] (toric1) at ($(0,0)+(-3.5,-1.4)$) {};
    \node [draw=black, circle, scale=0.5, thick, fill=black] (toric2) at ($(toric1)+(-1,{sqrt(3)})$) {};
    \node [draw=black, circle, scale=0.5, thick, fill=black] (toric3) at ($(toric1)+(-2,0)$) {};
    \node [draw=black, circle, scale=0.5, thick, fill=black] (toric4) at ($(toric1)+(-1,{3/5*sqrt(3)})$) {};
    \node [draw=black, circle, scale=0.5, thick, fill=black] (toric5) at ($(toric1)+(-1,{1/5*sqrt(3)})$) {};
    %Arrows
    \draw [ultra thick](toric1) -- (toric2);
    \draw [ultra thick](toric2) -- (toric3);
    \draw [ultra thick](toric3) -- (toric1);

%Quiver diagram
    % Nodes
    \node [draw=black, circle, inner sep=3.8pt, thick, fill=white] (su1) at (90:1.3) {$e$};
    \node [draw=black, circle, inner sep=1pt, thick, fill=white] (su5) at (162:1.3) {\scriptsize{$w_Y^4$}};
    \node [draw=black, circle, inner sep=1pt, thick, fill=white] (su4) at (234:1.3) {\scriptsize{$w_Y^3$}};
    \node [draw=black, circle, inner sep=1pt, thick, fill=white] (su3) at (306:1.3) {\scriptsize{$w_Y^2$}};
    \node [draw=black, circle, inner sep=1.8pt, thick, fill=white] (su2) at (18:1.3) {\scriptsize{$w_Y$}};
   % Arrows
   \draw [->>-] (su1) -- (su2);
   \draw [->>-] (su2) -- (su3);
   \draw [->>-] (su3) -- (su4);
   \draw [->>-] (su4) -- (su5);
   \draw [->>-] (su5) -- (su1);
   \draw [->-] (su1) -- (su4);
   \draw [->-] (su4) -- (su2);
   \draw [->-] (su2) -- (su5);
   \draw [->-] (su5) --  (su3);
   \draw [->-] (su3) -- (su1);

\node [] at (-4.5,-2) {(a)};
\node [] at (0,-2) {(b)};
\end{tikzpicture}
\caption{The toric diagram of $\CC^3/\ZZ_5$ (a) and the corresponding quiver diagram (b) are shown.}
\label{z5quiver}
\end{center}
\end{figure}

Let us move on to the case with $n=2$: ${\cal X}=\CC^3/\ZZ_5$.
In this case $w_X\neq w_Y=w_Z$ and the system has a flavor symmetry $SU(2)$.
The toric diagram and the quiver diagram are shown in Figure \ref{z5quiver}.
There are five sectors labeled by the wrapping numbers $w=w_Y^k$ ($k=0,1,2,3,4$).
The order of indices of these five sectors can be easily read off from
Table \ref{z5.tbl}.
\begin{table}[htb]
\caption{Wrapping numbers $w\in\Gamma$ and the corresponding monomials $F(X_I)$ are shown
for the orbifold ${\cal X}=\CC^3/\ZZ_5$. The shaded rows are irrelevant to the leading corrections.}
\label{z5.tbl}
\vspace{3mm}
\centering
\begin{tabular}{ccc}
\hline
\hline
$w$ & ${\cal O}(q^0)$ and ${\cal O}(q^N)$ & higher order \\
\hline
$e$         & $1$      & $XY^2$, ... \\
$w_Y$       & $Y$, $Z$ & $X^2$, ... \\
\sh $w_Y^2$ &\sh       &\sh $Y^2$, ... \\
$w_Y^3$     & $X$      &    $Y^3$, ... \\
\sh $w_Y^4$ &\sh       &\sh $XY$, ... \\
\hline
\end{tabular}
\end{table}
The sector with $w=e$ is the mesonic sector and sectors with $w=w_Y$ and $w=w_Y^3=w_X$
are baryonic sectors whose indices are ${\cal O}(q^N)$.
The other sectors with $w=w_Y^2$ and $w=w_Y^4$, which are shaded in
Table \ref{z5.tbl}, give indices of ${\cal O}(q^{2N})$
and we do not discuss them.

The expected relations are
\begin{align}
{\cal I}_0^{\rm gauge}
&={\cal I}^{\rm grav}+{\cal O}(q^{3N}),\nonumber\\
\frac{{\cal I}^{\rm gauge}_{\bm{b}_e-\bm{b}_{w_Y}}}{{\cal I}^{\rm gauge}_0}
&={\cal I}^{\rm D3}_{Y,0}+{\cal I}^{\rm D3}_{Z,0}+{\cal O}(q^{2N}),
\nonumber\\
\frac{{\cal I}^{\rm gauge}_{\bm{b}_e-\bm{b}_{w_Y^3}}}{{\cal I}^{\rm gauge}_0}
&={\cal I}^{\rm D3}_{X,0}+{\cal O}(q^{3N}).
\label{z5expected}
\end{align}
Let us confirm these relations for $N=2$.
For the mesonic sector with $w=e$
the result of localization on the gauge theory side is
\begin{align}
{\cal I}^{\rm gauge}_0
=&1+\left(u^5-\frac{\chi _3}{u}+\chi _5\right)q^5 -5\left( u \chi _2+1\right)q^6+\mathcal{O}(q^7).
\label{z5im}
\end{align}
The formula
for the Kaluza-Klein modes (\ref{igravdef}) gives
\begin{align}
{\cal I}^{\rm grav}=\cdots(\text{identical terms})\cdots+0q^6+\mathcal{O}(q^7).
\label{z5igrav}
\end{align}
(\ref{z5im})
and
(\ref{z5igrav}) are consistent with the first relation
in (\ref{z5expected}).

For the baryonic sector with $w=w_Y$
the result of localization with $N=2$
is
\begin{align}
\frac{{\cal I}^{\rm gauge}_{\bm{b}_e-\bm{b}_{w_Y}}}{{\cal I}^{\rm gauge}_0}
&=\chi _2q^2+ u^2 \chi _1q^3+ \chi^J_1\left(\chi _2-\frac{1}{u}\right)q^{\frac{7}{2}}+u^4q^4\nn\\
&\quad+\left(\chi _2^J(\chi_2-\frac{1}{u})-\frac{3}{u}-\chi _2\right)q^5+2 \chi _1^Ju^4q^{\frac{11}{2}}+\mathcal{O}(q^6),
\label{z5gaugeb1}
\end{align}
and on the gravity side we have
\begin{align}
{\cal I}^{\rm D3}_{Y,0}+{\cal I}^{\rm D3}_{Z,0}
&=\cdots(\text{identical terms})\cdots+ \chi _1^J u^4q^{\frac{11}{2}}+\mathcal{O}(q^6).
%=====================================
\label{z5gravityb1}
\end{align}
(\ref{z5gaugeb1}) and
(\ref{z5gravityb1}) satisfy the second relation in (\ref{z5expected}).

For the baryonic sector with $w=w_Y^3$
the result of localization with $N=2$
is
\begin{align}
\frac{{\cal I}^{\rm gauge}_{\bm{b}_e-\bm{b}_{w_Y^3}}}{{\cal I}^{\rm gauge}_0}
&=u^2q^2+ u^2 \chi _1^Jq^{\frac{7}{2}}+ u \chi _3q^4+u^2\left(\chi _2^J-1\right)q^5 +\left(\frac{\chi _2}{u^2}+\chi _6\right)q^6 \nonumber\\
&\quad+\left(u^2 \chi_3^J-u^2 \chi _1^J\right)q^{\frac{13}{2}} -\left(u^2 \chi _5+4 u \chi _3+2 \chi _1\right)q^7 +\mathcal{O}(q^{\frac{15}{2}}),
\label{z5b2gauge}
\end{align}
and on the gravity side we have
\begin{align}
{\cal I}^{\rm D3}_{X,0}
&=\cdots(\text{identical terms})\cdots-\left(u^2 \chi _5+u \chi _3\right)q^7+\mathcal{O}(q^{\frac{15}{2}}).
\label{z5b2gravity}
\end{align}
(\ref{z5b2gauge}) and (\ref{z5b2gravity}) satisfy the last relation in
(\ref{z5expected}).

%%%%%%%%%%%%%%%%%%%%%%%%%%%%%%%%%%%%%%%%%%%%%%%%%%%%%%%%%%%%%%%%
\subsubsection{$\CC^3/\ZZ_7$}
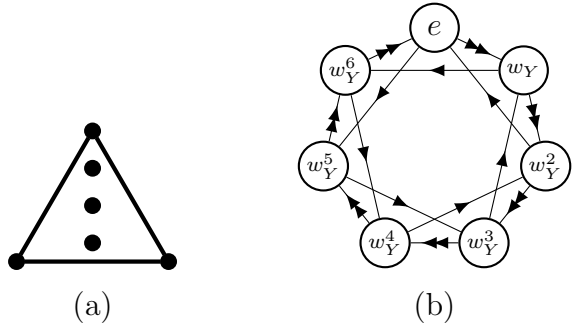
\begin{figure}[htbp]
\begin{center}
\begin{tikzpicture}
\tikzset{
  ->>-/.style={decoration={markings, mark=at position 0.55 with {\arrow[scale=1.5]{latex}}, mark=at position 0.79 with {\arrow[scale=1.5]{latex}}}, postaction={decorate}}
}
%Toric diagram
   %Nodes
    \node [draw=black, circle, scale=0.5, thick, fill=black] (toric1) at (-3.5,-1.6) {};
    \node [draw=black, circle, scale=0.5, thick, fill=black] (toric2) at ($(toric1)+(-1,{sqrt(3)})$) {};
    \node [draw=black, circle, scale=0.5, thick, fill=black] (toric3) at ($(toric1)+(-2,0)$) {};
    \node [draw=black, circle, scale=0.5, thick, fill=black] (toric4) at ($(toric1)+(-1,{5*sqrt(3)/7})$) {};
    \node [draw=black, circle, scale=0.5, thick, fill=black] (toric5) at ($(toric1)+(-1,{sqrt(3)/7})$) {};
    \node [draw=black, circle, scale=0.5, thick, fill=black] (toric6) at ($(toric1)+(-1,{3*sqrt(3)/7})$) {};
    %Arrows
    \draw [ultra thick](toric1) -- (toric2);
    \draw [ultra thick](toric2) -- (toric3);
    \draw [ultra thick](toric3) -- (toric1);

%Quiver diagram
    % Nodes
    \node [draw=black, circle, inner sep=3.8pt, thick, fill=white] (su1) at (90:1.5) {$e$};
    \node [draw=black, circle, inner sep=1.8pt, thick, fill=white] (su2) at (90-360/7:1.5) {\scriptsize{$w_Y$}};
    \node [draw=black, circle, inner sep=1pt, thick, fill=white] (su3) at (90-2*360/7:1.5) {\scriptsize{$w_Y^2$}};
    \node [draw=black, circle, inner sep=1pt, thick, fill=white] (su4) at (90-3*360/7:1.5) {\scriptsize{$w_Y^3$}};
    \node [draw=black, circle, inner sep=1pt, thick, fill=white] (su5) at (90-4*360/7:1.5) {\scriptsize{$w_Y^4$}};
    \node [draw=black, circle, inner sep=1pt, thick, fill=white] (su6) at (90-5*360/7:1.5) {\scriptsize{$w_Y^5$}};
    \node [draw=black, circle, inner sep=1pt, thick, fill=white] (su7) at (90-6*360/7:1.5) {\scriptsize{$w_Y^6$}};
   % Arrows
   %YZ
   \draw [->>-] (su1) -- (su2);
   \draw [->>-] (su2) -- (su3);
   \draw [->>-] (su3) -- (su4);
   \draw [->>-] (su4) -- (su5);
   \draw [->>-] (su5) -- (su6);
   \draw [->>-] (su6) -- (su7);
   \draw [->>-] (su7) -- (su1);
   %X
   \draw [->-] (su1) -- (su6);
   \draw [->-] (su6) -- (su4);
   \draw [->-] (su4) -- (su2);
   \draw [->-] (su2) -- (su7);
   \draw [->-] (su7) -- (su5);
   \draw [->-] (su5) -- (su3);
   \draw [->-] (su3) -- (su1);

\node [] at (-4.5,-2.2) {(a)};
\node [] at (0,-2.2) {(b)};
\end{tikzpicture}
\caption{The toric diagram of $\CC^3/\ZZ_7$ (a) and the corresponding quiver diagram (b) are shown.}
\label{z7quiver2}
\end{center}
\end{figure}

For $n=3$ the orbifold is ${\cal X}=\CC^3/\ZZ_7$
with the orbifold group generated by
$\diag(\omega_7^{-2},\omega_7,\omega_7)$.
The toric diagram and the quiver diagram are
shown in Figure \ref{z7quiver2}.
There are seven sectors specified by the wrapping numbers $w=w_Y^k$ ($k=0,1,2,\ldots,6$).
(See Table \ref{z7.tbl}.)
\begin{table}[htb]
\caption{Wrapping numbers $w\in\Gamma$ and the corresponding monomials $F(X_I)$ are shown
for the orbifold ${\cal X}=\CC^3/\ZZ_7$. The shaded rows are irrelevant to the leading corrections.}
\label{z7.tbl}
\vspace{3mm}
\centering
\begin{tabular}{ccc}
\hline
\hline
$w$ & ${\cal O}(q^0)$ and ${\cal O}(q^N)$ & higher order \\
\hline
    $e$     & $1$      & $XY^2$, ... \\
    $w_Y$   & $Y$, $Z$ & $X^3$, ... \\
\sh $w_Y^2$ &\sh       &\sh $Y^2$, ... \\
\sh $w_Y^3$ &\sh       &\sh $X^2$, ... \\
\sh $w_Y^4$ &\sh       &\sh $X^2Y$, ... \\
    $w_Y^5$ & $X$      & $X^2Y^2$, ... \\
\sh $w_Y^6$ &\sh       &\sh $XY$, ... \\
\hline
\end{tabular}
\end{table}

We are interested in the mesonic sector with $w=e$ and two baryonic sectors
with $w=w_Y$ and $w=w_Y^5$.
The other sectors shaded in Table \ref{z7.tbl} give
higher order corrections which we do not discuss here.
The expected relations are
\begin{align}
{\cal I}^{\rm gauge}_0
&={\cal I}^{\rm grav}+{\cal O}(q^{3N}),\nonumber\\
\frac{{\cal I}^{\rm gauge}_{\bm{b}_e-\bm{b}_{w_Y}}}{{\cal I}^{\rm gauge}_0}
&={\cal I}^{\rm D3}_{Y,0}+{\cal I}^{\rm D3}_{Z,0}+{\cal O}(q^{3N}),\nonumber\\
\frac{{\cal I}^{\rm gauge}_{\bm{b}_e-\bm{b}_{w_Y^5}}}{{\cal I}^{\rm gauge}_0}
&={\cal I}^{\rm D3}_{X,0}+{\cal O}(q^{4N}).
\label{z7expected}
\end{align}
Let us confirm these relations for $N=2$.

For the mesonic sector with $w=e$ we have on the gauge theory side
\begin{align}
\mathcal{I}^{\rm gauge}_0
&=1-7\left( u \chi _2+1\right)q^6 +\mathcal{O}(q^7).
\end{align}
and on the gravity side
\begin{align}
\mathcal{I}^{\text{grav}}&=1+0q^6+\mathcal{O}(q^7).
\end{align}
These are consistent with the first relation in (\ref{z7expected}).
For the baryonic sector with $w=w_Y$
the index on the gauge theory side is
\begin{align}
\frac{{\cal I}^{\rm gauge}_{\bm{b}_e-\bm{b}_{w_Y}}}{{\cal I}^{\rm gauge}_0}
&=\chi _2q^2
+\chi^J_1 \left(\chi _2 -\frac{1}{u}\right)q^{\frac{7}{2}}
+ u^3 \chi _1q^4
+\left(\chi _2^J\chi_2-\frac{\chi _2^J}{u}-\chi_2-\frac{3}{u}\right)q^5\nn\\
&\quad+u^6q^6+\chi_3^J\left(-\frac{1}{u}+\chi_2\right)q^{\frac{13}{2}}-6u^3\chi_1q^7+\mathcal{O}(q^{\frac{15}{2}}),
\end{align}
and on the gravity side we have
\begin{align}
\mathcal{I}^{\mathrm{D}3}_{Y,0}+\mathcal{I}^{\mathrm{D}3}_{Z,0}
&=\cdots(\text{identical terms})\cdots-4u^3\chi_1q^7+\mathcal{O}(q^{\frac{15}{2}}).
\end{align}
These are consistent with the second equation in (\ref{z7expected}).
For the baryonic sector with $w=w_Y^5$
on the gauge theory side we have
\begin{align}
\frac{{\cal I}^{\rm gauge}_{\bm{b}_e-\bm{b}_{w_Y^5}}}{{\cal I}^{\rm gauge}_0}
&=u^2q^2
+ u^2 \chi _1^Jq^{\frac{7}{2}}
+ \left(u^2 \chi _2^J-u^2\right)q^5+ u \chi _5q^6
+u^2\left(-\chi_1^J+\chi_3^J\right)q^{\frac{13}{2}}\nonumber\\
&\quad
+u^2\left(1-\chi_2^J+\chi_4^J+u\chi_2\right)q^8+\mathcal{O}(q^{\frac{17}{2}}),
\end{align}
and on the gravity side we have
\begin{align}
\mathcal{I}^{\mathrm{D}3}_{X,0}
&=\cdots(\text{identical terms})\cdots+u^2\left(-\chi_2^J+\chi_4^J\right)q^8+\mathcal{O}(q^9).
\end{align}
These are consistent with the third equation in (\ref{z7expected}).

%%%%%%%%%%%%%%%%%%%%%%%%%%%%%%%%%%%%%%%%%%%%%%%%%%%%
\subsection{The other $\CC^3/\ZZ_7$}
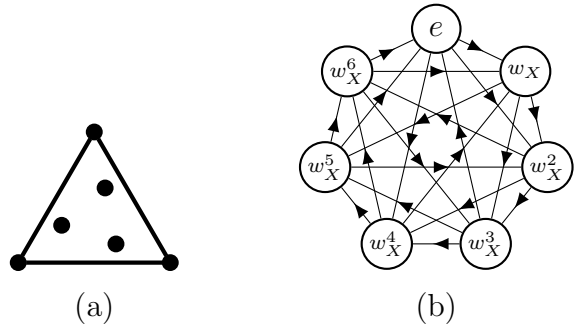
\begin{figure}[htbp]
\begin{center}
\begin{tikzpicture}
\tikzset{
  ->-/.style={decoration={markings, mark=at position 0.58 with {\arrow[scale=1.5]{latex}}},
              postaction={decorate}},
  ->>-/.style={decoration={markings, mark=at position 0.55 with {\arrow[scale=1.5]{latex}}, mark=at position 0.79 with {\arrow[scale=1.5]{latex}}},
              postaction={decorate}},
  ->>>-/.style={decoration={markings, mark=at position 0.47 with {\arrow[scale=2]{latex}}, mark=at position 0.59 with {\arrow[scale=2]{latex}}, mark=at position 0.71 with {\arrow[scale=2]{latex}}}, postaction={decorate}}
}
%Toric diagram
   %Nodes
    \node [draw=black, circle, scale=0.5, thick, fill=black] (toric1) at (-3.5,-1.6) {};
    \node [draw=black, circle, scale=0.5, thick, fill=black] (toric2) at ($(toric1)+(-1,{sqrt(3)})$) {};
    \node [draw=black, circle, scale=0.5, thick, fill=black] (toric3) at ($(toric1)+(-2,0)$) {};
    \node [draw=black, circle, scale=0.5, thick, fill=black] (toric4) at ($(toric1)+({-5/7},{sqrt(3)/7})$) {};
    \node [draw=black, circle, scale=0.5, thick, fill=black] (toric5) at ($(toric1)+({-10/7},{2*sqrt(3)/7})$) {};
    \node [draw=black, circle, scale=0.5, thick, fill=black] (toric6) at ($(toric1)+({-6/7},{4*sqrt(3)/7})$) {};
    %Arrows
    \draw [ultra thick](toric1) -- (toric2);
    \draw [ultra thick](toric2) -- (toric3);
    \draw [ultra thick](toric3) -- (toric1);

%Quiver diagram
    % Nodes
    \node [draw=black, circle, inner sep=3.8pt, thick, fill=white] (su1) at (90:1.5) {$e$};
    \node [draw=black, circle, inner sep=1.8pt, thick, fill=white] (su2) at (90-360/7:1.5) {\scriptsize{$w_X$}};
    \node [draw=black, circle, inner sep=1pt, thick, fill=white] (su3) at (90-2*360/7:1.5) {\scriptsize{$w_X^2$}};
    \node [draw=black, circle, inner sep=1pt, thick, fill=white] (su4) at (90-3*360/7:1.5) {\scriptsize{$w_X^3$}};
    \node [draw=black, circle, inner sep=1pt, thick, fill=white] (su5) at (90-4*360/7:1.5) {\scriptsize{$w_X^4$}};
    \node [draw=black, circle, inner sep=1pt, thick, fill=white] (su6) at (90-5*360/7:1.5) {\scriptsize{$w_X^5$}};
    \node [draw=black, circle, inner sep=1pt, thick, fill=white] (su7) at (90-6*360/7:1.5) {\scriptsize{$w_X^6$}};
   % Arrows
   %X
   \draw [->-] (su1) -- (su2);
   \draw [->-] (su2) -- (su3);
   \draw [->-] (su3) -- (su4);
   \draw [->-] (su4) -- (su5);
   \draw [->-] (su5) -- (su6);
   \draw [->-] (su6) -- (su7);
   \draw [->-] (su7) -- (su1);
   %Y
   \draw [->-] (su1) -- (su3);
   \draw [->-] (su3) -- (su5);
   \draw [->-] (su5) -- (su7);
   \draw [->-] (su7) -- (su2);
   \draw [->-] (su2) -- (su4);
   \draw [->-] (su4) -- (su6);
   \draw [->-] (su6) -- (su1);
   %Z
   \draw [->-] (su1) -- (su5);
   \draw [->-] (su5) -- (su2);
   \draw [->-] (su2) -- (su6);
   \draw [->-] (su6) -- (su3);
   \draw [->-] (su3) -- (su7);
   \draw [->-] (su7) -- (su4);
   \draw [->-] (su4) -- (su1);

\node [] at (-4.5,-2.2) {(a)};
\node [] at (0,-2.2) {(b)};
\end{tikzpicture}

\caption{The toric diagram of $\CC^3/\ZZ_7$ (a) and the corresponding quiver diagram (b) are shown.}
\label{z7quiver1}
\end{center}
\end{figure}

There is the other $\ZZ_7$ orbifold different from the one studied in the last subsection.
Its orbifold group $\ZZ_7$ is generated by
$\diag(\omega_7,\omega_7^2,\omega_7^4)$
and the dual group $\Gamma$ is characterized by the relations
\begin{align}
w_X^2=w_Y,\quad
w_Y^2=w_Z,\quad
w_Z^2=w_X,\quad
w_Xw_Yw_Z=e.
\end{align}
The corresponding toric diagram and the quiver diagram are shown in Figure \ref{z7quiver1}.
Among the seven sectors specified by
$w=w_X^k$ ($k=0,1,2,\ldots,6$)
we are interested in one mesonic sector $w=e$
and three baryonic sectors $w=w_X$, $w=w_Y=w_X^2$,
and $w=w_Z=w_X^4$.
(See Table \ref{z72.tbl})
\begin{table}[htb]
\caption{Wrapping numbers $w\in\Gamma$ and the corresponding monomials $F(X_I)$ are shown
for the orbifold ${\cal X}=\CC^3/\ZZ_7$. The shaded rows are irrelevant to the leading corrections.}
\label{z72.tbl}
\vspace{3mm}
\centering
\begin{tabular}{ccc}
\hline
\hline
 $w$ & ${\cal O}(q^0)$ and ${\cal O}(q^N)$ & higher order \\
\hline
    $e$     & $1$ & $XYZ$, ... \\
    $w_X$   & $X$ & $Z^2$, ... \\
    $w_X^2$ & $Y$ & $X^2$, ... \\
\sh $w_X^3$ &\sh  &\sh $XY$, ... \\
    $w_X^4$ & $Z$ & $Y^2$, ... \\
\sh $w_X^5$ &\sh  &\sh $XZ$, ... \\
\sh $w_X^6$ &\sh  &\sh $YZ$, ... \\
\hline
\end{tabular}
\end{table}
The expected relations for these sectors are
\begin{align}
{\cal I}^{\rm gauge}_0&={\cal I}^{\rm grav}+{\cal O}(q^{3N}),\nonumber\\
\frac{{\cal I}^{\rm gauge}_{\bm{b}_e-\bm{b}_{w_X}}}{{\cal I}^{\rm gauge}_0}
&={\cal I}^{\rm D3}_{X,0}+{\cal O}(q^{2N}),\nonumber\\
\frac{{\cal I}^{\rm gauge}_{\bm{b}_e-\bm{b}_{w_X^2}}}{{\cal I}^{\rm gauge}_0}
&={\cal I}^{\rm D3}_{Y,0}+{\cal O}(q^{2N}),\nonumber\\
\frac{{\cal I}^{\rm gauge}_{\bm{b}_e-\bm{b}_{w_X^4}}}{{\cal I}^{\rm gauge}_0}
&={\cal I}^{\rm D3}_{Z,0}+{\cal O}(q^{2N}).
\label{z72expected}
\end{align}
Let us confirm these equations for $N=2$.

The calculation on the gauge theory side gives the following index for the mesonic sector $w=e$:
\begin{align}
\mathcal{I}^{\rm gauge}_0&=1-14 q^6+\mathcal{O}(q^7).
\end{align}
On the gravity side we have the following Kaluza-Klein index.
\begin{align}
\mathcal{I}^{\text{grav}}&=1+0q^6+\mathcal{O}(q^7).
\end{align}
These two are consistent with the first relation in (\ref{z72expected}).

The latter three equations in (\ref{z72expected})
for the baryonic sectors are related by
cyclic permutations of $X$, $Y$, and $Z$, and we explicitly check only the second in (\ref{z72expected}).
On the gauge theory side the index is calculated as
\begin{align}
\frac{{\cal I}^{\rm gauge}_{\bm{b}_e-\bm{b}_{w_X}}}{{\cal I}^{\rm gauge}_0}
&= u^2q^2+\frac{ u}{v^2}q^3
+ u^2 \chi _1^Jq^{\frac{7}{2}}
+ \left(\frac{v}{u}+\frac{1}{v^4}\right)q^4
\nonumber\\&\quad
+\left(u^2 \chi _2^J-u^2 +\frac{1}{u^2v}+\frac{v^4}{u^3}\right)q^5+\mathcal{O}(q^{\frac{11}{2}}),
\end{align}
and on the gravity side we obtain
\begin{align}
\mathcal{I}^{\mathrm{D}3}_{X,0}
&=\cdots(\text{identical terms})\cdots+\left(u^2 \chi _2^J-u^2+\frac{1}{u^2v}+\frac{v^4}{u^3}+\frac{1}{u v^6}\right)q^5 +\mathcal{O}(q^6).\end{align}
These two results are consistent with the second relation in (\ref{z72expected}).

%%%%%%%%%%%%%%%%%%%%%%%%%%%%%%%%%%%%%%%%%%%%%%%%%%%%
%%%%%%%%%%%%%%%%%%%%%%%%%%%%%%%%%%%%%%%%%%%%%%%%%%%%
\subsection{$\CC\times\CC^2/\ZZ_n$ ($n=2,3,4,\ldots$)}
Let us consider the series of orbifolds
${\cal X}=\CC\times\CC/\ZZ_n$ ($n=2,3,4,\ldots$).
We assume that $\ZZ_n$ acts on $Y$ and $Z$ as
\begin{align}
(X,Y,Z)\rightarrow(X,\omega_nY,\omega_n^{-1}Z).
\end{align}
The dual group $\Gamma$ is characterized by the relations
\begin{align}
w_X=w_Yw_Z=w_Y^n=e.
\end{align}
If $n=2$, the relation $w_Y=w_Z$ holds and the system has
an $SU(2)$ flavor symmetry while if $n\geq 3$ $w_Y\neq w_Z$ and there is no
continuous flavor symmetry.
An important feature of these orbifolds is the existence of the fixed locus
$Y=Z=0$,
and we need to take account of the contribution of the tensor multiplets.

Although these systems preserve ${\cal N}=2$ supersymmetry
the indices we calculate in this section are
not ${\cal N}=2$ indices
because of the difference in the treatment of the
trace parts in the ${\cal N}=1$ vector multiplets and adjoint chiral multiplets.
We include the trace contribution for the adjoint chiral multiplets
while we do not for the vector multiplets.

\subsubsection{$\CC\times\CC^2/\ZZ_2$}
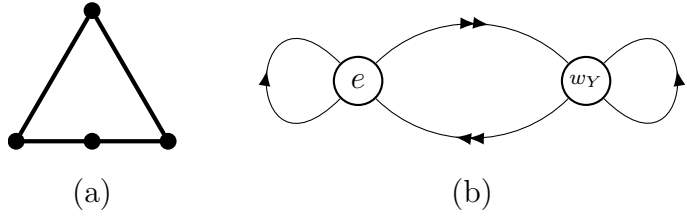
\begin{figure}[htb]
\centering
\begin{tikzpicture}
\tikzset{
  ->>-/.style={decoration={markings, mark=at position 0.52 with {\arrow[scale=1.5]{latex}}, mark=at position 0.58 with {\arrow[scale=1.5]{latex}}}, postaction={decorate}}
}
\clip  (-6.1,-1.8) rectangle (2.8,3);
%Toric diagram
   %Nodes
    \node [draw=black, circle, scale=0.5, thick, fill=black] (toric1) at (-4,-0.8) {};
    \node [draw=black, circle, scale=0.5, thick, fill=black] (toric2) at ($(toric1)+(-1,{sqrt(3)})$) {};
    \node [draw=black, circle, scale=0.5, thick, fill=black] (toric3) at ($(toric1)+(-2,0)$) {};
    %Arrows
    \draw [ultra thick](toric1) -- (toric2);
    \draw [ultra thick](toric2) -- (toric3);
    \draw [ultra thick](toric3) -- (toric1);
    \node [draw=black, circle, scale=0.5, thick, fill=black] (toric5) at ($(toric1)+(-1,0)$) {};

   %Quiver diagram
    % Nodes
    \node [draw=black, circle, inner sep=3.8pt, thick, fill=white] (su1) at (180:1.5) {$e$};
    \node [draw=black, circle, inner sep=1.8pt, thick, fill=white] (su2) at (0:1.5) {\scriptsize{$w_Y$}};
   % Arrows
   %X
   \draw [->-] (su1) to [out=225,in=135,loop,looseness=10] (su1);
   \draw [->-] (su2) to [out=315,in=45,loop,looseness=10] (su2);
   %Y,Z
   \draw [->>-] (su1) to [out=45,in=135] (su2);
   \draw [->>-] (su2) to [out=225,in=315] (su1);

\node [] at (-5,-1.5) {(a)};
\node [] at (0,-1.5) {(b)};
\end{tikzpicture}
\caption{The toric diagram of $\CC\times\CC^2/\ZZ_2$ (a) and the corresponding quiver diagram (b) are shown.}\label{c2z2c.eps}
\end{figure}

Let us first consider $n=2$ case: $\CC\times\CC/\ZZ_2$.
The toric diagram and the quiver diagram are shown in Figure \ref{c2z2c.eps}.
There are two sectors specified by $w=e$ and $w=w_Y$ (Table \ref{cc2z2.tbl}).
\begin{table}[htb]
\caption{Wrapping numbers $w\in\Gamma$ and the corresponding monomials $F(X_I)$ are shown
for the orbifold ${\cal X}=\CC\times\CC^2/\ZZ_2$.}
\label{cc2z2.tbl}
\vspace{3mm}
\centering
\begin{tabular}{ccc}
\hline
\hline
$w$ & ${\cal O}(q^0)$ and ${\cal O}(q^N)$ & higher order \\
\hline
$e$ & $1$, $X$ & $X^2$, ... \\
$w_Y$ & $Y$, $Z$ & $XY$, ... \\
\hline
\end{tabular}
\end{table}
The expected relations are
\begin{align}
{\cal I}^{\rm gauge}_0
&={\cal I}^{\rm grav}{\cal I}_X^{\rm tensor}(1+2{\cal I}_{X,0}^{\rm D3})+{\cal O}(q^{2N}),\nonumber\\
\frac{{\cal I}^{\rm gauge}_{\bm{b}_e-\bm{b}_{w_Y}}}{{\cal I}^{\rm gauge}_0}
&={\cal I}_{Y,0}^{\rm D3}+{\cal I}_{Z,0}^{\rm D3}+{\cal O}(q^{2N}).
\label{cc2z2expexted}
\end{align}

We can numerically confirm the relations in (\ref{cc2z2expexted}) for $N=2$.
For the mesonic sector with $w=e$
the calculation on the gauge theory side gives
\begin{align}
{\cal I}^{\rm gauge}_0
&=1
+2uq
+\left(-\frac{1}{u}+5u^2+\chi_2\right)q^2
+\left(-2+8u^3+2u\chi_2\right)q^3
+2u^2\chi^J_1q^{\frac{7}{2}}
\nonumber\\
&\quad +\left(\frac{1}{u^2}-7u+14u^4-\frac{2}{u}\chi_2+3u^2\chi_2+2\chi_4\right)q^4
+\left(4u^3\chi^J_1+2u\chi^J_1\chi_2\right)q^{\frac{9}{2}}
\nonumber\\
&\quad +\left(-14u^2+20u^5+2u^2\chi^J_2-6\chi_2+4u^3\chi_2+2u\chi_4\right)q^5
\nonumber\\
&\quad +\left(\frac{2}{u^2}\chi^J_1-2u\chi^J_1+10u^4\chi^J_1+6u^2\chi^J_1\chi_2+2\chi^J_1\chi_4 \right)q^{\frac{11}{2}}
\nonumber\\
&\quad +\left(-\frac{1}{u^3}-25u^3+30u^6+4u^3\chi^J_2+\frac{1}{u^2}\chi_2-13u\chi_2+5u^4\chi_2-\frac{2}{u}\chi_4 \right.
\nonumber\\
&\hspace{2cm}
  +3u^2\chi_4+2\chi_6 \bigg) q^6
+{\cal O}(q^{\frac{13}{2}}).
\end{align}
On the gravity side we obtain
\begin{align}
&{\cal I}^{\rm grav}
{\cal I}_X^{\rm tensor}
(1+2{\cal I}^{\rm D3}_{X,0})
=\cdots(\mbox{identical terms})\cdots
\nonumber\\
&\quad+\bigg(-1-\frac{4}{u^3}-28u^3+29u^6+4u^3\chi^J_2-16u\chi_2+4u^4\chi_2
-\frac{5}{u}\chi_4
\nonumber\\
&\hspace{2cm}
+2u^2\chi_4+\chi_6\bigg)q^6
+{\cal O}(q^{\frac{13}{2}}).
\end{align}
These results are consistent with the first equation in (\ref{cc2z2expexted}).

For the baryonic sector with $w=w_Y$ the localization on the gauge theory side gives
\begin{align}
\frac{{\cal I}^{\rm gauge}_{\bm{b}_e-\bm{b}_{w_Y}}}{{\cal I}^{\rm gauge}_0}
&=\chi_2q^2
+\left(1-u\chi_2\right)q^3
+\left(-\frac{1}{u}\chi^J_1+\chi^J_1\chi_2\right)q^{\frac{7}{2}}
+\left(-\frac{2}{u^2}-u-u^2\chi_2\right)q^4
\nonumber\\
&\quad +2\chi^J_1q^{\frac{9}{2}}
+\left(-u^2-\frac{1}{u}\chi^J_2-2\chi_2+u^3\chi_2+\chi^J_2\chi_2\right)q^5
+{\cal O}(q^{\frac{11}{2}}),
\end{align}
and the analysis on the gravity side gives
\begin{align}
{\cal I}^{\rm D3}_{Y,0}+{\cal I}^{\rm D3}_{Z,0}
&=\cdots(\mbox{identical terms})\cdots
\nonumber\\
&\quad +\left(-u^2-\frac{1}{u}\chi^J_2-2\chi_2-u^3\chi_2+\chi^J_2\chi_2\right)q^5
+{\cal O}(q^{\frac{11}{2}}).
\end{align}
These two results are consistent with the second equation in (\ref{cc2z2expexted}).

%%%%%%%%%%%%%%%%%%%%%%%%%%%%%%%%%%%%%%%%%%%%%%
\subsubsection{$\CC\times\CC^2/\ZZ_3$}
\begin{figure}[htb]
\centering
\begin{tikzpicture}
   \clip (-6.8,-2.3) rectangle (2.8,3);
%Toric diagram
   %Nodes
    \node [draw=black, circle, scale=0.5, thick, fill=black] (toric1) at (-3.5,-1.2) {};
    \node [draw=black, circle, scale=0.5, thick, fill=black] (toric2) at ($(toric1)+(-1,{sqrt(3)})$) {};
    \node [draw=black, circle, scale=0.5, thick, fill=black] (toric3) at ($(toric1)+(-2,0)$) {};
    %Arrows
    \draw [ultra thick](toric1) -- (toric2);
    \draw [ultra thick](toric2) -- (toric3);
    \draw [ultra thick](toric3) -- (toric1);
    \node [draw=black, circle, scale=0.5, thick, fill=black] (toric4) at ($(toric1)+(-2/3,0)$) {};
    \node [draw=black, circle, scale=0.5, thick, fill=black] (toric5) at ($(toric1)+(-4/3,0)$) {};

   %Quiver diagram
    % Nodes
    \node [draw=black, circle, inner sep=3.8pt, thick, fill=white] (su1) at (90:1.5) {$e$};
    \node [draw=black, circle, inner sep=1.8pt, thick, fill=white] (su2) at (330:1.5) {\scriptsize{$w_Y$}};
    \node [draw=black, circle, inner sep=1pt, thick, fill=white] (su3) at (210:1.5) {\scriptsize{$w_Y^2$}};
   % Arrows
   %X
   \draw [->-] (su1) to [out=135,in=45,loop,looseness=10] (su1);
   \draw [->-] (su2) to [out=15,in=285,loop,looseness=10] (su2);
   \draw [->-] (su3) to [out=255,in=165,loop,looseness=10] (su3);
   %Y
   \draw [->-] (su1) to [out=315,in=105] (su2);
   \draw [->-] (su2) to [out=195,in=345] (su3);
   \draw [->-] (su3) to [out=75,in=225] (su1);
   %Z
   \draw [->-] (su2) to [out=135,in=285] (su1);
   \draw [->-] (su1) to [out=255,in=45] (su3);
   \draw [->-] (su3) to [out=15,in=165] (su2);

\node [] at (-4.5,-2) {(a)};
\node [] at (0,-2) {(b)};
\end{tikzpicture}
\caption{The toric diagram of $\CC\times\CC^2/\ZZ_3$ (a) and the corresponding quiver diagram (b) are shown.}\label{c2znxc.eps}
\end{figure}
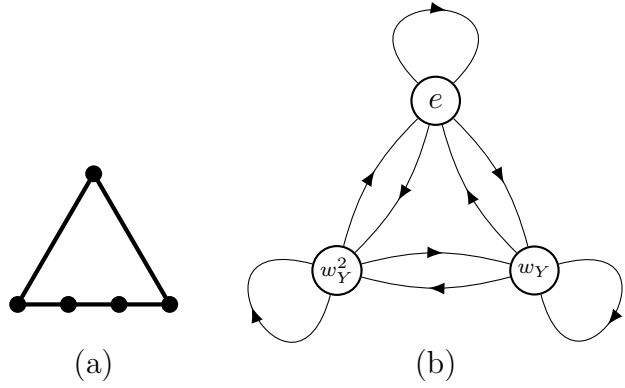

Let us consider the $n=3$ case with the orbifold ${\cal X}=\CC\times\CC^2/\ZZ_3$.
The toric diagram and the quiver diagram are shown in Figure \ref{c2znxc.eps}.
There are three sectors specified by $w=e,w_Y,w_Y^2$.
The sector with $w=e$ is the mesonic sector and
the ones with $w=w_Y$ and $w=w_Y^2$ are baryonic sectors.
See Table \ref{C2Z2Ca}.
\begin{table}[htb]
\caption{Wrapping numbers $w\in\Gamma$ and the corresponding monomials $F(X_I)$ are shown
for the orbifold ${\cal X}=\CC\times\CC^2/\ZZ_3$.}
\label{C2Z2Ca}
\vspace{3mm}
\centering
\begin{tabular}{ccc}
\hline
\hline
$w$ & ${\cal O}(q^0)$ and ${\cal O}(q^N)$ & higher order \\
\hline
$e$ & $1$, $X$ & $X^2$, ... \\
$w_Y$ & $Y$      & $XY$, ... \\
$w_Y^2$ & $Z$ & $XZ$, ... \\
\hline
\end{tabular}
\end{table}

The expected relations are
\begin{align}
{\cal I}^{\rm gauge}_0
&={\cal I}^{\rm grav}({\cal I}_X^{\rm tensor})^2(1+3{\cal I}_{X,0}^{\rm D3})+{\cal O}(q^{2N}),\nonumber\\
\frac{{\cal I}^{\rm gauge}_{\bm{b}_e-\bm{b}_{w_Y}}}{{\cal I}^{\rm gauge}_0}
&={\cal I}_{Y,0}^{\rm D3}+{\cal O}(q^{2N}),\nonumber\\
\frac{{\cal I}^{\rm gauge}_{\bm{b}_e-\bm{b}_{w_Y^2}}}{{\cal I}^{\rm gauge}_0}
&={\cal I}_{Z,0}^{\rm D3}+{\cal O}(q^{2N}).
\label{eq114}
\end{align}

We numerically confirm these relations for $N=2$.
For the mesonic sector
the calculation on the gauge theory side gives
\begin{align}
{\cal I}^{\rm gauge}_0&=
1
+3uq
+9u^2q^2
+\left(19u^3+\tfrac{1}{v^3}+\tfrac{v^3}{u^3}\right)q^3
+3u^2\chi^J_1q^{\frac{7}{2}}
\nonumber\\
&\quad+\left(-6u+39u^4+\frac{3u}{v^3}+\frac{3v^3}{u^2}\right)q^4
+\left(3\chi^J_1+9u^3\chi^J_1\right)q^{\frac{9}{2}}
\nonumber\\
&\quad+\left(-\frac{6}{u}-21u^2+69u^5+\frac{6u^2}{v^3}+\frac{6v^3}{u}+3u^2\chi^J_2\right)q^5
\nonumber\\
&\quad+\left(\frac{3}{u^2}\chi^J_1+9u\chi^J_1+27u^4\chi^J_1+\frac{3u}{v^3}\chi^J_1+\frac{3v^3}{u^2}\chi^J_1\right)q^{\frac{11}{2}}
\nonumber\\
&\quad+\bigg(-21+\frac{1}{u^3}-57u^3+119u^6+\frac{2}{v^6}-\frac{6}{v^3}+\frac{10u^3}{v^3}+10v^3-\frac{6v^3}{u^3}
\nonumber\\&\hspace{2cm}
    +\frac{2v^6}{u^6}+9u^3\chi^J_2\bigg)q^6
+{\cal O}(q^{\frac{13}{2}}).
\end{align}
On the gravity side we obtain
\begin{align}
&
{\cal I}^{\rm grav}({\cal I}^{\rm tensor})^2(1+3{\cal I}^{\rm D3}_{X=0})
=\cdots(\mbox{identical terms})\cdots
\nonumber\\
&\quad +\left(-27-\frac{5}{u^3}-63u^3+116u^6+9u^3\chi^J_2+\frac{6}{u}\chi_1-10u^2\chi_1-6\chi_3+10u^3\chi_3\right.
\nonumber\\
&\hspace{2cm}\left.-\frac{2}{u}\chi_4+2\chi_6\right)q^6
+{\cal O}(q^{\frac{13}{2}})
\end{align}
These two results are consistent with the
first equation in (\ref{eq114}).

Two baryonic sectors with $w=w_Y$ and $w=w_Y^2$
are related by the Weyl reflection swapping $Y$ and $Z$,
and hence we check only the relation for $w=w_Y$ explicitly.
The localization on the gauge theory side gives
\begin{align}
\frac{{\cal I}^{\rm gauge}_{\bm{b}_e-\bm{b}_{w_Y}}}{{\cal I}^{\rm gauge}_0}
&=\frac{v^2}{u^2}q^2
+\left(\frac{1}{uv}-\frac{v^2}{u}\right)q^3
+\frac{v^2}{u^2}\chi^J_1q^{\frac{7}{2}}
+\left(\frac{1}{v^4}-v^2\right)q^4
\nonumber\\
&\quad +\left(-\frac{2u}{v^4}-\frac{1}{u^2v}-\frac{3u}{v}-\frac{v^2}{u^2}+uv^2+\frac{v^2}{u^2}\chi^J_2\right)q^5
+{\cal O}(q^{\frac{11}{2}})
\end{align}
and on the gravity side we obtain
\begin{align}
{\cal I}_{Y,0}^{\rm D3}&=\cdots(\mbox{identical terms})\cdots
\nonumber\\
&\quad+\left(\frac{u}{v^7}-\frac{1}{u^2v}-\frac{u}{v}-\frac{v^2}{u^2}+\frac{v^2}{u^2}\chi^J_2\right)q^5
+{\cal O}(q^{\frac{11}{2}}).
\end{align}
These results confirm the second equation in (\ref{eq114}).

%%%%%%%%%%%%%%%%%%%%%%%%%%%%%%%%%%%%%%%%%%%%%%%%%%%%
\subsection{$\mathbb{C}^3/\mathbb{Z}_4$}
\begin{figure}[htbp]
\begin{center}
\begin{center}
\begin{tikzpicture}
%Toric diagram
   %Nodes
    \node [draw=black, circle, scale=0.5, thick, fill=black] (toric1) at (-3,-1.5) {};
    \node [draw=black, circle, scale=0.5, thick, fill=black] (toric2) at ($(toric1)+(-1,{sqrt(3)})$) {};
    \node [draw=black, circle, scale=0.5, thick, fill=black] (toric3) at ($(toric1)+(-2,0)$) {};
    \node [draw=black, circle, scale=0.5, thick, fill=black] (toric4) at ($(toric1)+(-1,{sqrt(3)/2})$) {};
    %Arrows
    \draw [ultra thick](toric1) -- (toric2);
    \draw [ultra thick](toric2) -- (toric3);
    \draw [ultra thick](toric3) -- (toric1);
    \node [draw=black, circle, scale=0.5, thick, fill=black] (toric5) at ($(toric1)+(-1,0)$) {};

   %Quiver diagram
    % Nodes
    \node [draw=black, circle, inner sep=3.8pt, thick, fill=white] (su1) at (90:1.5) {$e$};
    \node [draw=black, circle, inner sep=1.8pt, thick, fill=white] (su2) at (90-360/4:1.5) {\scriptsize{$w_Y$}};
    \node [draw=black, circle, inner sep=1pt, thick, fill=white] (su3) at (90-2*360/4:1.5) {\scriptsize{$w_Y^2$}};
    \node [draw=black, circle, inner sep=1pt, thick, fill=white] (su4) at (90-3*360/4:1.5) {\scriptsize{$w_Y^3$}};
   % Arrows
   %X(0, 2.5) .. controls (5,-0.7) .. (10, 2.5)
   \draw [->-] (su1) to [out=250,in=110] (su3);
   \draw [->-] (su3) to [out=70,in=290] (su1);
   \draw [->-] (su2) to [out=160,in=20] (su4);
   \draw [->-] (su4) to [out=340,in=200] (su2);
   %Y,Z
   \draw [->>-] (su1) -- (su2);
   \draw [->>-] (su2) -- (su3);
   \draw [->>-] (su3) -- (su4);
   \draw [->>-] (su4) -- (su1);

\node [] at (-4,-2.4) {(a)};
\node [] at (0,-2.4) {(b)};
\end{tikzpicture}
\end{center}
\caption{The toric diagram of $\CC^3/\ZZ_4$ (a) and the corresponding quiver diagram (b) are shown.}
\label{quiverz4}
\end{center}
\end{figure}
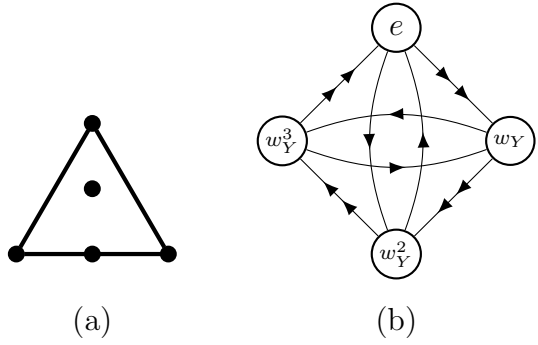

Let us consider the orbifold group $\ZZ_4$ generated by
$\diag(\omega_4^{-2},\omega_4,\omega_4)$.
The toric diagram and the quiver diagram are shown in Figure
\ref{quiverz4}.

The dual group $\Gamma=\ZZ_4$ is characterized by
the relations
\begin{align}
w_Y=w_Z,\quad
w_Y^2=w_X,\quad
w_X^2=e.
\end{align}
There are four sectors specified by $w=w_Y^k$ ($k=0,1,2,3$).
(Table \ref{z44.tbl})
\begin{table}[htb]
\caption{Wrapping numbers $w\in\Gamma$ and the corresponding monomials $F(X_I)$ are shown
for the orbifold ${\cal X}=\CC^3/\ZZ_4$. The shaded row is irrelevant to the leading corrections.}
\label{z44.tbl}
\vspace{3mm}
\centering
\begin{tabular}{ccc}
\hline
\hline
$w$ & ${\cal O}(q^0)$ and ${\cal O}(q^N)$ & higher order \\
\hline
    $e$     & $1$      & $X^2$, ... \\
    $w_Y$   & $Y$, $Z$ & $X^2Y$, ... \\
    $w_Y^2$ & $X$      & $YZ$, ... \\
\sh $w_Y^3$ &\sh       &\sh $XY$, ... \\
\hline
\end{tabular}
\end{table}

The expected relations are
\begin{align}
{\cal I}^{\rm gauge}
&={\cal I}^{\rm grav}{\cal I}_X^{\rm tensor}+{\cal O}(q^{2N}),\nonumber\\
\frac{{\cal I}^{\rm gauge}_{\bm{b}_e-\bm{b}_{w_Y}}}{{\cal I}^{\rm gauge}_0}
&={\cal I}_{Y,0}^{\rm D3}+{\cal I}_{Z,0}^{\rm D3}+{\cal O}(q^{3N}),\nonumber\\
\frac{{\cal I}^{\rm gauge}_{\bm{b}_e-\bm{b}_{w_Y^2}}}{{\cal I}^{\rm gauge}_0}
&={\cal I}_{X,0}^{\rm D3}+{\cal O}(q^{2N}).
\label{z44expected}
\end{align}
Let us confirm these relations for $N=2$.

For the mesonic sector with $w=e$,
the calculation on the gauge theory side gives
\begin{align}
\mathcal{I}^{\rm gauge}_0
&=1
+2u^2q^2
+\left(5u^4-\frac{\chi_2}{u}+\chi_4\right)q^4
-4u^2q^5+\mathcal{O}(q^{\frac{11}{2}}).
\end{align}
On the gravity side we obtain
\begin{align}
{\cal I}^{\rm grav} {\cal I}_X^{\rm tensor}
&=1+2u^2q^2+\left(5u^4-\frac{\chi_2}{u}+\chi_4\right)q^4+0q^5+\mathcal{O}(q^6).
\end{align}
These two results are consistent with the first relation in
(\ref{z44expected}).

For the baryonic sector with $w=w_Y$ the gauge theory calculation gives
\begin{align}
\frac{{\cal I}^{\rm gauge}_{\bm{b}_e-\bm{b}_{w_Y}}}{{\cal I}^{\rm gauge}_0}
&=\chi_2 q^2
 +\chi _1^J\left(\chi_2 -\frac{1}{u}\right)q^{\frac{7}{2}}
+ \left(u-\chi_2 u^2\right)q^4
\nonumber\\
&\quad +\left(\chi_2  \left(\chi _2^J-1\right)-\frac{\chi _2^J}{u}-\frac{3}{u}\right)q^5
+2  u \chi _1^J q^{\frac{11}{2}}
\nonumber\\
&\quad-\left(\chi_2 \left(u^4+\frac{1}{u^2}\right)+u^3-\frac{1}{u^3}\right)q^6
+ \chi _3^J \left(\chi_2 -\frac{1}{u}\right)q^{\frac{13}{2}}
\nonumber\\
&\quad+ u\left(\chi_2 u-3\right)q^7
   +\mathcal{O}(q^{\frac{15}{2}}),
\end{align}
and the analysis on the gravity side gives
\begin{align}
{\cal I}_{Y,0}^{\rm D3}+{\cal I}_{Z,0}^{\rm D3}
&=\cdots(\mbox{identical terms})\cdots- u \left(\chi_2 u+1\right)q^7+\mathcal{O}(q^{\frac{15}{2}}).
\end{align}
These two results are consistent with the second relation in (\ref{z44expected}).

For the baryonic sector with $w=w_Y^2$
the calculation on the gauge theory side gives
\begin{align}
\frac{{\cal I}^{\rm gauge}_{\bm{b}_e-\bm{b}_{w_Y^2}}}{{\cal I}^{\rm gauge}_0}
&=u^2q^2+u\chi_2q^3+u^2\chi^J_1q^{\frac{7}{2}}+\left(\frac{1}{u^2}+\chi_4\right)q^4\nonumber\\
&\quad +u^2\left(\chi^J_2-1\right)q^5+\mathcal{O}(q^{\frac{11}{2}}),
\end{align}
and the analysis on the gravity side gives
\begin{align}
{\cal I}_{X,0}^{\rm D3}
&=\cdots(\mbox{identical terms})\cdots
+\left(u^2\left(\chi^J_2-1\right)+\frac{1}{u}\chi_6+\frac{\chi_2}{u^3}\right)q^5+\mathcal{O}(q^{\frac{11}{2}}).
\end{align}
These results are consistent with the third relation
in (\ref{z44expected}).

%%%%%%%%%%%%%%%%%%%%%%%%%%%%%%%%%%%%%%%%%%%%%%%%%%%%%%%%%%%%%%%%%%%%%%%
\subsection{$\CC^3/(\ZZ_2\times\ZZ_2)$}
\begin{figure}[htbp]
\begin{center}
\begin{tikzpicture}
%Toric diagram
   %Nodes
    \node [draw=black, circle, scale=0.5, thick, fill=black] (toric1) at (-3,-1.5) {};
    \node [draw=black, circle, scale=0.5, thick, fill=black] (toric2) at ($(toric1)+(-1,{sqrt(3)})$) {};
    \node [draw=black, circle, scale=0.5, thick, fill=black] (toric3) at ($(toric1)+(-2,0)$) {};
    %Arrows
    \draw [ultra thick](toric1) -- (toric2);
    \draw [ultra thick](toric2) -- (toric3);
    \draw [ultra thick](toric3) -- (toric1);
    \node [draw=black, circle, scale=0.5, thick, fill=black] (toric4) at ($(toric1)+(-1,0)$) {};
    \node [draw=black, circle, scale=0.5, thick, fill=black] (toric5) at ($(toric1)+(-1/2,{sqrt(3)/2})$) {};
    \node [draw=black, circle, scale=0.5, thick, fill=black] (toric6) at ($(toric1)+(-3/2,{sqrt(3)/2})$) {};

%Quiver diagram
    % Nodes
    \node [draw=black, circle, inner sep=3.8pt, thick, fill=white] (su1) at (90:1.5) {$e$};
    \node [draw=black, circle, inner sep=1.8pt, thick, fill=white] (su2) at (90-360/4:1.5) {\scriptsize{$w_X$}};
    \node [draw=black, circle, inner sep=1.8pt, thick, fill=white] (su3) at (90-2*360/4:1.5) {\scriptsize{$w_Y$}};
    \node [draw=black, circle, inner sep=1.8pt, thick, fill=white] (su4) at (90-3*360/4:1.5) {\scriptsize{$w_Z$}};
   % Arrows
   %X
   \draw [->-] (su1) to [out=300,in=150] (su2);
   \draw [->-] (su2) to [out=120,in=330] (su1);
   \draw [->-] (su3) to [out=120,in=330] (su4);
   \draw [->-] (su4) to [out=300,in=150] (su3);
   %Y
   \draw [->-] (su1) to [out=255,in=105] (su3);
   \draw [->-] (su3) to [out=75,in=285] (su1);
   \draw [->-] (su2) to [out=165,in=15] (su4);
   \draw [->-] (su4) to [out=345,in=195] (su2);
   %Z
   \draw [->-] (su1) to [out=210,in=60] (su4);
   \draw [->-] (su4) to [out=30,in=240] (su1);
   \draw [->-] (su2) to [out=210,in=60] (su3);
   \draw [->-] (su3) to [out=30,in=240] (su2);

\node [] at (-4,-2.4) {(a)};
\node [] at (0,-2.4) {(b)};
\end{tikzpicture}
\caption{The toric diagram of $\CC^3/(\ZZ_2\times\ZZ_2)$ (a) and the corresponding quiver diagram (b) are shown.}
\label{z2z2quiver}
\end{center}
\end{figure}
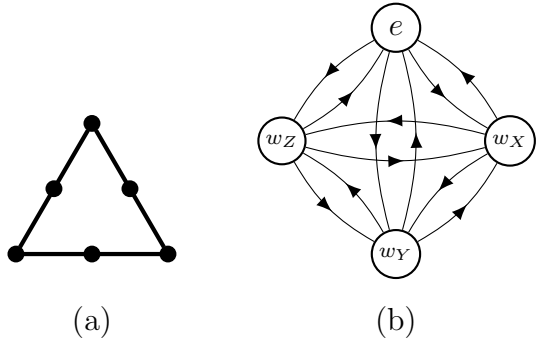

Finally, let us consider the orbifold with
$\wt\Gamma=\ZZ_2\times\ZZ_2$ generated by $\diag(1,-1,-1)$ and $(-1,-1,1)$
The corresponding toric diagram and the quiver diagram are shown in Figure \ref{z2z2quiver}.
The dual group $\Gamma=\ZZ_2\times\ZZ_2$ is characterized by the following relations.
\begin{align}
w_X^2=w_Y^2=w_Z^2=w_Xw_Yw_Z=e.
\end{align}
There are four sectors specified by $w=e,w_X,w_Y,w_Z$.
(Table \ref{z2z2.tbl})
\begin{table}[htb]
\caption{Wrapping numbers $w\in\Gamma$ and the corresponding monomials $F(X_I)$ are shown
for the orbifold ${\cal X}=\CC^3/(\ZZ_2\times\ZZ_2)$.}
\label{z2z2.tbl}
\vspace{3mm}
\centering
\begin{tabular}{ccc}
\hline
\hline
$w$ & ${\cal O}(q^0)$ and ${\cal O}(q^N)$ & higher order \\
\hline
$e$ & $1$ & $X^2$, ... \\
$w_X$ & $X$ & $YZ$, ... \\
$w_Y$ & $Y$ & $XZ$, ... \\
$w_Z$ & $Z$ & $XY$, ... \\
\hline
\end{tabular}
\end{table}

The expected relations are
\begin{align}
{\cal I}^{\rm gauge}_0
&={\cal I}^{\rm grav}
{\cal I}^{\rm tensor}_X
{\cal I}^{\rm tensor}_Y
{\cal I}^{\rm tensor}_Z+{\cal O}(q^{2N}),\nonumber\\
\frac{{\cal I}^{\rm gauge}_{\bm{b}_e-\bm{b}_{w_X}}}{{\cal I}^{\rm gauge}_0}
&={\cal I}^{\rm D3}_{X,0}+{\cal O}(q^{2N}),\nonumber\\
\frac{{\cal I}^{\rm gauge}_{\bm{b}_e-\bm{b}_{w_Y}}}{{\cal I}^{\rm gauge}_0}
&={\cal I}^{\rm D3}_{Y,0}+{\cal O}(q^{2N}),\nonumber\\
\frac{{\cal I}^{\rm gauge}_{\bm{b}_e-\bm{b}_{w_Z}}}{{\cal I}^{\rm gauge}_0}
&={\cal I}^{\rm D3}_{Z,0}+{\cal O}(q^{2N}).
\label{z2z2expected}
\end{align}
Let us conform these relations for $N=2$.

We first consider the mesonic sector with $w=e$.
On the gauge theory side the index for $N=2$ is
\begin{align}
\mathcal{I}^{\rm gauge}_0
&=1+2\left( \chi _{(2,0)}- \chi _{(0,1)}\right)q^2
+ \left(4 \chi _{(0,2)}+\chi _{(1,0)}-5 \chi _{(2,1)}+5 \chi _{(4,0)}\right)q^4\nn\\
&\quad+4\left(\chi _{(0,1)}- \chi_{(2,0)}\right)q^5 +\mathcal{O}(q^{\frac{11}{2}}),
\end{align}
and on the gravity side we obtain
\begin{align}
{\cal I}^{\rm grav}
{\cal I}^{\rm tensor}_X
{\cal I}^{\rm tensor}_Y
{\cal I}^{\rm tensor}_Z
=&\cdots(\text{identical terms})\cdots+0q^5+\mathcal{O}(q^6).
\end{align}
These results are consistent with the first relation in (\ref{z2z2expected}).

The three baryonic sectors are related among them by permutations among $X$, $Y$, and $Z$.
We check explicitly the second equation in (\ref{z2z2expected})
associated with the baryonic sector with $w=w_X$.
On the gauge theory side we obtain
\begin{align}
\frac{{\cal I}^{\rm gauge}_{\bm{b}_e-\bm{b}_{w_X}}}{{\cal I}^{\rm gauge}_0}
&=u^2q^2
+q^3
+ u^2 \chi _1^Jq^{\frac{7}{2}}
+ \left(-\frac{u^2}{v^2}+\frac{1}{u^2}-v^2\right)q^4
\nonumber\\&\quad
+ \left(u^2 \chi _2^J-u^2\right)q^5
+\mathcal{O}(q^{\frac{11}{2}}),
\end{align}
and on the gravity side we obtain
\begin{align}
{\cal I}^{\rm D3}_{X,0}
&=\cdots(\text{identical terms})\cdots+\left(u^2 \chi _2^J+\frac{1}{u^4}-u^2\right)q^5 +\mathcal{O}(q^6).
\end{align}
These two results are consistent with the second relation in (\ref{z2z2expected}).

%%%%%%%%%%%%%%%%%%%%%%%%%%%%%%%%%%%%%%%%%%%%%%%%%%%%%%%%%%%%%%%%%%%%%%%%%
\section{Summary and discussions}\label{discussion.sec}
We investigated the finite $N$ corrections
to the superconformal index for quiver gauge theories
realized on D3-branes put in abelian orbifolds
${\cal X}=\CC^3/\wt\Gamma$.
We focused on the leading corrections starting from $q^N$
and gave a prescription to calculate them
on the gravity side as the contribution of wrapped D3-branes.
Based on the analysis in \cite{Arai:2019xmp} for
S-fold theories we assumed that
the corrections are reproduced
as the contributions of D3-branes
wrapped over three particular three-cycles,
$X=0$, $Y=0$, and $Z=0$
in the internal space ${\cal Y}=\bm{S}^5/\wt\Gamma$.
The D3-brane worldvolume of such a configuration
is expressed by using a monomial function $F(X,Y,Z)$
as $F(X,Y,Z)=0$.

We established the relation between the D3-brane wrapping number
and the baryonic charges.
We gave a simple prescription to obtain
baryonic charges
from the function $F$
supplemented by holonomy variables.
We compared
indices calculated on the both sides of
the AdS/CFT correspondence
for each sector specified by
the wrapping number of D3-branes.

In this paper we focused on the
leading finite $N$ corrections,
and did not study sub-leading corrections
starting at $q^{kN}$ with $k\geq2$
depending on the sector we considered.
As far as we have checked for
examples we found the complete agreement up to
higher order terms of sub-leading corrections.
This fact strongly suggests that our method is correct at least for
configurations with a single D3-brane.

There are many open questions.
An interesting and important task
is the test of our method for the
sub-leading corrections.
In particular, analytic expressions for the exact Schur index is known for a few theories with finite $N$ \cite{Bourdier:2015wda,Bourdier:2015sga} and it would be very nice
if we could reproduce the complete index on the AdS side.
Our assumption claims that they should
be reproduced by D3-brane configurations
including two or more wrapped D3-branes.
Naively, the index for such configurations
would be calculated as the index of
non-abelian gauge theories
realized on the coincident wrapped D3-branes.
In addition, if the configuration contains branes wrapped on
two different cycles open strings connecting
them may contribute to the index.
Further investigation is required to
establish the method to calculate these contributions.

In the examples ${\cal X}=\CC\times\CC/\ZZ_n$
we found that D3-branes wrapped on
a topologically trivial cycle
give non-trivial contribution
to the index.
These branes are
often referred to as giant gravitons,
and such non-trivial contribution
to the index
had been also found in ${\cal N}=4$ SYM
in \cite{Arai:2019xmp}.
On such branes a ``tachyonic'' degree of freedom lives
and there exists a $q^{-1}$ term in the corresponding
single-particle index.
We formally deal with the plethystic exponential
of this term
by the relation
\begin{align}
\Pexp(q^{-1})=-q\Pexp(q).
\end{align}
The negative sign appearing in this relation
correctly reproduces the finite $N$ correction
due to ``the absence of giant gravitons.''
At present, unfortunately, we have no physical understanding
of this mysterious
agreement.

One potential way to justify our method
rigorously is to
use localization.
Let ${\cal M}$ be the configuration space of
D3-branes in the internal space ${\cal Y}$.
The three configurations, $X=0$, $Y=0$, and $Z=0$
are fixed points of a rotation
of ${\cal M}$ induced by an isometry of ${\cal Y}$.
If we can extend ${\cal M}$ so that the path integral of the fields
on D3-branes are treated as the integration over
the extended space,
and we can localize the integral to the points corresponding to
the branes wrapped on the three cycles
by using a localization theorem,
then it would give a justification of the
prescription we proposed in this paper.

In this paper we studied abelian orbifolds.
There seems no obstacles to extend
the method to non-abelian orbifolds
and toric Sasaki-Einstein manifolds.
In particular, for the latter class of manifolds
the dual gauge theories are generically strongly coupled and the agreement
of the superconformal index is highly non-trivial.
There exists systematic prescription \cite{Hanany:2005ss}
to obtain the field contents
of quiver gauge theory from the
toric data of the dual internal space ${\cal Y}$,
and it is in principle possible
to calculate the index by using the localization formula.
On the gravity side, as a naive analogy of the prescription for orbifolds
we expect that the finite $N$ correction may be reproduced as the contribution
of D3-branes wrapped on three-cycles in ${\cal Y}$.
Although it may not be straightforward
to calculate the single-particle
indices for excitations on such three-cycles,
once we obtain them we can easily test the proposal.
We hope that we can come back to this problem in near future.

\section*{Acknowledgments}
We would like to thank D.~Yokoyama
for valuable discussions and comments.

\appendix

\section{Results for $N=3$}\label{n3.sec}

In this appendix we show the results
for $N=3$.
On the gravity side the calculation finishes instantly on a typical laptop computer.
while it takes much longer time than the $N=2$ case
to carry out the calculation on the gauge theory side.
For some cases we could not obtain ${\cal I}^{\rm gauge}_w$
up to sufficient order to find disagreement between ${\cal I}^{\rm AdS}_w$
and ${\cal I}^{\rm gauge}_w$.
In the following we show the results for examples
in which we could successfully finish the calculation up to the
sufficient order.

\subsection{$\CC^3/\ZZ_3$}
The relation (\ref{z3mesonic}) is confirmed for $N=3$ by comparing the following results.
\begin{align}
{\cal I}^{\rm gauge}_0
&=1+\left(1-\chi_{(1,1)}+\chi_{(3,0)}\right)q^3\nonumber\\
&\quad+\left(1+\chi_{(0,3)}-\chi_{(1,1)}+2\chi_{(3,0)}-2\chi_{(4,1)}+2\chi_{(6,0)}\right)q^6\nonumber\\
&\quad +\left(2 \chi_{(0,3)}-2 \chi_{(1,1)}-2 \chi_{(1,4)}-3 \chi_{(2,2)}-\chi_{(3,0)}+2 \chi_{(3,3)}-2 \chi_{(4,1)}\right.\nonumber\\
&\hspace{2cm} \left.-3 \chi_{(7,1)}+3 \chi_{(9,0)}-2\right)q^9+\mathcal{O}(q^{\frac{21}{2}}),\\
\mathcal{I}^{\mathrm{grav}}
&=\cdots(\text{identical terms})\cdots\nonumber\\
&\quad+\left(1+2\chi_{(0,3)}-2\chi_{(1,1)}-2\chi_{(1,4)}+2\chi_{(3,0)}+2\chi_{(3,3)}-2\chi_{(4,1)}\right.\nonumber\\
&\hspace{2cm} \left. +3\chi_{(6,0)}-3\chi_{(7,1)}+3\chi_{(9,0)}\right)q^9+\mathcal{O}(q^{12}).
\end{align}
The relation (\ref{b1z3}) is confirmed for $N=3$ by comparing the following results.
\begin{align}
&\frac{{\cal I}^{\rm gauge}_{\bm{b}_e-\bm{b}_{w_X}}}{{\cal I}^{\rm gauge}_0}\nonumber\\
&=\chi_{(3,0)}q^3
+ (\chi_{(3,0)}-\chi_{(1,1)}) \chi _1^Jq^{\frac{9}{2}}
\nonumber\\
&\quad+\left((-\chi_{(1,1)}+\chi_{(3,0)}+1) \chi _2^J-2 \chi_{(0,3)}-\chi_{(3,0)}\right)q^6
\nonumber\\
   &\quad+ \left((-\chi_{(0,3)}+2 \chi_{(1,1)}+\chi_{(2,2)}-\chi_{(3,0)}+1) \chi _1^J+(-\chi_{(1,1)}+\chi_{(3,0)}+1) \chi _3^J\right)q^{\frac{15}{2}}\nonumber\\
   &\quad+ \left((2 \chi_{(1,1)}-2 \chi_{(3,0)}-1) \chi
   _2^J+(-\chi_{(1,1)}+\chi_{(3,0)}+1) \chi _4^J\right.\nonumber\\
   &\hspace{2cm}\left.+2 \chi_{(0,3)}+\chi_{(1,4)}+\chi_{(2,2)}-2 \chi_{(3,0)}-\chi_{(4,1)}-\chi_{(5,2)}+5\right)q^9\nonumber\\
   &\quad+ \left((\chi_{(0,3)}-4 \chi_{(1,1)}+\chi_{(1,4)}-3 \chi_{(2,2)}-3 \chi_{(3,0)}-2\chi_{(3,3)}+\chi_{(4,1)}+\chi_{(6,0)} \right.\nonumber\\
   &\hspace{2cm}\left.+\chi_{(7,1)}-1)\chi _1^J+(\chi_{(1,1)}-\chi_{(3,0)}-1) \left(\chi
   _3^J-\chi _5^J\right)\right)q^{\frac{21}{2}}\nonumber\\
   &\quad+
   \left((\chi_{(0,3)}-2 \chi_{(1,1)}-\chi_{(2,2)}+4 \chi_{(3,0)}-\chi_{(3,3)}+4 \chi_{(4,1)}+\chi_{(5,2)}-\chi_{(6,0)}-1) \chi _2^J\right.\nonumber\\
   &\hspace{2cm}-\chi_{(1,1)} \chi _6^J+(\chi_{(1,1)}-\chi_{(3,0)}-1) \chi _4^J+\chi_{(3,0)} \chi _6^J+5
   \chi_{(0,3)}+6 \chi_{(1,1)}\nonumber\\
   &\hspace{2cm}+\chi_{(1,4)}+6 \chi_{(2,2)}+12 \chi_{(3,0)}+5 \chi_{(3,3)}+15 \chi_{(4,1)}+8
   \chi_{(5,2)}+4 \chi_{(6,0)}\nonumber\\
   &\hspace{2cm}\left.+\chi_{(6,3)}+\chi_{(7,1)}+\chi_{(9,0)}-\chi_{(10,1)}+\chi_6^J-10\right)q^{12}+\mathcal{O}(q^{\frac{25}{2}}),
\end{align}
\begin{align}
&{\cal I}^{\rm D3}_{X,0}+{\cal I}^{\rm D3}_{Y,0}+{\cal I}^{\rm D3}_{Z,0}
\nonumber\\
&=\cdots(\text{identical terms})\cdots\nonumber\\
&\quad+ \left((\chi_{(0,3)}-2 \chi_{(1,1)}-\chi_{(2,2)}+4 \chi_{(3,0)}-\chi_{(3,3)}+4 \chi_{(4,1)}+\chi_{(5,2)}\right.\nonumber\\
&\hspace{2cm}-\chi_{(6,0)}-1) \chi _2^J-\chi_{(1,1)} \chi _6^J+(\chi_{(1,1)}-\chi_{(3,0)}-1) \chi _4^J+\chi_{(3,0)} \chi_6^J\nonumber\\
&\hspace{2cm}+\chi_{(1,1)}-\chi_{(1,4)}+\chi_{(2,2)}+4 \chi_{(3,0)}+8 \chi_{(4,1)}+3 \chi_{(5,2)}+2 \chi_{(6,0)}\nonumber\\
&\hspace{2cm}\left.+\chi_{(6,3)}-\chi_{(7,1)}-\chi_{(9,0)}-\chi_{(10,1)}+\chi _6^J-5\right)q^{12}+\mathcal{O}(q^{\frac{25}{2}}).
\end{align}

%%%%%%%%%%%%%%%%%%%%%%%%%%%%%%%%%%%%%%%%%%%%%%
\subsection{$\CC^3/\ZZ_5$}
The first relation in (\ref{z5expected}) is confirmed for $N=3$
by comparing the following results.
\begin{align}
{\cal I}^{\rm gauge}_0
&=1+\left(u^5-\frac{\chi _3}{u}+\chi _5\right)q^5 -5\left(2+u\chi_2+u^2\chi_4\right)q^9+\mathcal{O}(q^{\frac{19}{2}}),
\\
\mathcal{I}^{\mathrm{grav}}
&=\cdots(\text{identical terms})\cdots+0q^9+\mathcal{O}(q^{10}).
\end{align}
The second relation in (\ref{z5expected}) is confirmed for $N=3$
by comparing the following results.
\begin{align}
\frac{{\cal I}^{\rm gauge}_{\bm{b}_e-\bm{b}_{w_Y}}}{{\cal I}^{\rm gauge}_0}
&=\chi _3q^3 + u^2 \chi _2q^4+\chi^J_1\left(\chi _3 -\frac{\chi _1}{u}\right)q^{\frac{9}{2}} +u^4 \chi _1q^5 - u \chi _1^Jq^{\frac{11}{2}}\nn\\
&\quad+\left(-\frac{\chi _1 \chi _2^J}{u}+\chi _3 \chi_2^J+u^6-\frac{\chi _1}{u}-\chi _3\right)q^6 +\left(-2 u^2 \chi _2-u\right)q^7 \nn\\
&\quad+\left(2 u^6 \chi _1^J+\frac{2 \chi _1^{J}\chi_1}{u}-\frac{\chi _1 \chi _3^J}{u}-\chi _3 \chi_1^J+\chi _3^J\chi_3\right)q^{\frac{15}{2}} +\mathcal{O}(q^{8})
\end{align}
\begin{align}
{\cal I}^{\rm D3}_{Y,0}+{\cal I}^{\rm D3}_{Z,0}
&=\cdots(\text{identical terms})\cdots
\nonumber\\&\quad
+\left(u^6 \chi _1^J+\frac{2 \chi _1^{J}\chi_1}{u}-\frac{\chi _1 \chi _3^J}{u}-\chi _3 \chi_1^J+\chi _3^J\chi_3\right)q^{\frac{15}{2}} \nn\\
&\quad+\mathcal{O}(q^{8})
\end{align}
For the third relation of (\ref{z5expected}) we could not finish the calculation
on the gauge theory side.

%%%%%%%%%%%%%%%%%%%%%%%%%%%%%%%%%%%%%%%%%%%%%%

\subsection{$\CC\times\CC^2/\ZZ_2$}
The first relation in (\ref{cc2z2expexted}) is confirmed for $N=3$ by comparing the following results.
\begin{align}
\mathcal{I}^{\text{gauge}}_0
&=1+2  uq+ \left(5 u^2-\frac{1}{u}+\chi _2\right)q^2+ \left(10 u^3+2 u \chi _2-2\right)q^3\nn\\
&\quad+ \left(18 u^4+5 u^2 \chi _2+\frac{1}{u^2}-\frac{2 \chi _2}{u}-5 u+2\chi _4\right)q^4\nn\\
&\quad+2  u^3 \chi _1^Jq^{\frac{9}{2}}+ \left(30 u^5+8 u^3 \chi _2-12 u^2+4 u \chi _4+\frac{2}{u}-4 \chi _2\right)q^5\nn\\
&\quad+ \left(4 u^4\chi _1^J+2 u^2 \chi_2 \chi _1^J\right)q^{\frac{11}{2}}\nn\\
&\quad+ \left(2 u^3 \chi _2^J+49 u^6+14 u^4 \chi _2-24 u^3-\frac{2}{u^3}+\frac{2 \chi _2}{u^2}+8 u^2 \chi _4\right.\nn\\
&\hspace{2cm}\left.-12 u \chi _2-\frac{3 \chi _4}{u}+3\chi _6+3\right)q^6\nn\\
&\quad+ \left(10 u^5 \chi _1^J+6 u^3 \chi _2 \chi _1^J-2 u^2 \chi _1^J+2 u \chi _4 \chi _1^J+\frac{2 \chi _1^J}{u}\right)q^{\frac{13}{2}}\nn\\
&\quad+ \left(4 u^4\chi _2^J+74 u^7+20 u^5 \chi _2-44 u^4+12 u^3 \chi _4-26 u^2 \chi _2\right.\nn\\
&\hspace{2cm}\left.-\frac{6}{u^2}+\frac{2 \chi _2}{u}+4 u \chi _6+4 u-8 \chi _4\right)q^7\nn\\
&\quad+ \left(20 u^6\chi _1^J+14 u^4 \chi _2 \chi _1^J-6 u^3 \chi _1^J+2 u^3 \chi _3^J+\frac{2 \chi _2 \chi _1^J}{u^2}\right.\nn\\
&\hspace{2cm}+8 u^2 \chi _4 \chi _1^J+2 u \chi _2 \chi _1^J+2 \chi _6 \chi_1^J+6 \chi _1^J\bigg)q^{\frac{15}{2}}\nn\\
&\quad+ \left(10 u^5 \chi _2^J+2 u^3 \chi _2^{J+1}-2 u^2 \chi _2^J-2 \chi _2^J\chi_2+110 u^8+30 u^6 \chi _2\right.\nn\\
&\hspace{2cm}\left.-76 u^5+19 u^4 \chi_4+\frac{3}{u^4}-47 u^3 \chi _2-\frac{2 \chi _2}{u^3}+\frac{2 \chi _4}{u^2}+8 u^2 \chi _6\right.\nn\\
&\hspace{2cm}\left.+7 u^2-22 u \chi _4-\frac{4 \chi _6}{u}-\frac{13}{u}+\chi _2+4 \chi_8\right)q^8+\mathcal{O}(q^{\frac{17}{2}}),
\end{align}
\begin{align}
&{\cal I}^{\rm grav}
{\cal I}_X^{\rm tensor}
(1+2{\cal I}^{\rm D3}_{X,0})=\cdots(\text{identical terms})\cdots\nn\\
&\quad+ \left(10 u^5 \chi _2^J+2 u^3 \chi _2^{J+1}-2 u^2 \chi _2^J-2 \chi _2^J\chi_2+109 u^8+29 u^6 \chi _2\right.\nn\\
&\hspace{2cm}\left.-79 u^5+18 u^4 \chi_4+\frac{2}{u^4}-50 u^3 \chi _2-\frac{5 \chi _2}{u^3}+\frac{\chi _4}{u^2}+7 u^2 \chi _6\right.\nn\\
&\hspace{2cm}\left.+6 u^2-25 u \chi _4-\frac{7 \chi _6}{u}-\frac{16}{u}+3 \chi_8\right)q^8+\mathcal{O}(q^{\frac{17}{2}}).
\end{align}
The second relation in (\ref{cc2z2expexted}) is confirmed for $N=3$ by comparing the following results.
\begin{align}
\frac{{\cal I}^{\rm gauge}_{\bm{b}_e-\bm{b}_{w_Y}}}{{\cal I}^{\rm gauge}_0}
&= \chi _3q^3+ \left(\chi _1-u \chi _3\right)q^4+ \left(\chi _3 \chi _1^J-\frac{\chi _1^J\chi_1}{u}\right)q^{\frac{9}{2}}- \left(u^2 \left(\chi _3\right)+\frac{\chi_1}{u^2}\right)q^5\nn\\
&\quad+\chi _1^J\chi_1q^{\frac{11}{2}}+ \left(-\frac{\chi _1 \chi _2^J}{u}+\chi _3 \chi _2^J-2 u^2 \chi _1-\frac{\chi _1}{u}-\chi _3\right)q^6\nn\\
&\quad+\left(-u^2 \chi _3 \chi _1^J+\frac{\chi _1^J\chi_1}{u^2}+3 u \chi _1^J\chi_1\right)q^{\frac{13}{2}}\nonumber\\
&\quad+ \left(u^4 \chi _3-u^3 \chi _1-\frac{\chi _3}{u^2}-2 \chi _1\right)q^7\nn\\
&\quad+\left(-2 u^3 \chi _3 \chi _1^J+u^2 \chi _1^J\chi_1-\frac{\chi _1 \chi _3^J}{u}+\chi _3^J\chi_3\right)q^{\frac{15}{2}}+\mathcal{O}(q^8),
\end{align}
\begin{align}
{\cal I}^{\rm D3}_{Y,0}+{\cal I}^{\rm D3}_{Z,0}
&=\cdots(\text{identical terms})\cdots+ \left(u^2 \chi _1^J\chi_1-\frac{\chi _1 \chi _3^J}{u}+\chi _3^J\chi_3\right)q^{\frac{15}{2}}
\nonumber\\
&\quad+\mathcal{O}(q^8).
\end{align}

%%%%%%%%%%%%%%%%%%%%%%%%%%%%%%%%%%%%%%%%%%%%%%
\subsection{$\CC\times\CC^2/\ZZ_3$}
The first relation in (\ref{eq114}) is confirmed for $N=3$ by comparing the following results.
\begin{align}
&{\cal I}^{\rm gauge}_0\nonumber\\
&=1+3uq+9  u^2q^2+ \left(-\frac{\chi_1}{u}+\chi_3+22 u^3\right)q^3\nonumber\\
&\quad+ \left(3 \chi_3 u-3 \chi_1+48 u^4\right)q^4+3  u^3 \chi _1^Jq^{\frac{9}{2}}\nonumber\\
&\quad+ 3u \left( 3 u\chi_3+32 u^4-2u-3 \chi_1\right)q^5+3  \left(3 u^4+u\right) \chi _1^Jq^{\frac{11}{2}}\nonumber\\
&\quad+ \left(3 u^3 \chi _2^J+19 \chi_3 u^3-19
   \chi_1 u^2-\frac{2 \chi_4}{u}+2 \chi_6+182 u^6-21 u^3+\frac{1}{u^3}-6\right)q^6\nonumber\\
   &\quad+3  \chi _1^J \left(\chi_3
   u^2-\chi_1 u+9 u^5+3 u^2+\frac{1}{u}\right)q^{\frac{13}{2}}\nonumber\\
   &\quad+ \bigg(9 u^4 \chi _2^J+\chi_3 \left(39 u^4-6 u\right)+\chi_1 \left(6-39 u^3\right)+6 \chi_6u\nonumber\\
   &\hspace{2cm}\left.-6 \chi_4+324 u^7-63 u^4-\frac{3}{u^2}-27 u\right)q^7\nonumber\\
   &\quad+3  \left(\chi _1^J \left(\chi_3 \left(4 u^3+1\right)-\chi_1 \left(4
   u^2+\frac{1}{u}\right)+22 u^6+8 u^3+5 +\frac{1}{u^3}\right)\right.\nonumber\\
   &\hspace{2cm}+u^3 \chi _3^J\bigg)q^{\frac{15}{2}}\nonumber\\
   &\quad+3  \left(9 u^5 \chi _2^J+5 \chi_6 u^2-5 \chi_4 u+\chi_3
   \left(23 u^5-9 u^2-\frac{1}{u}\right)+185 u^8\right.\nonumber\\
   &\hspace{2cm}\left.-51 u^5-24 u^2-\frac{6}{u}+\chi_1 \left(-23 u^4+9 u+\frac{2}{u^2}\right)\right)q^8+\mathcal{O}(q^{\frac{17}{2}}),
\end{align}
\begin{align}
&{\cal I}^{\rm grav}({\cal I}_X^{\rm tensor})^{2}(1+3{\cal I}_{X,0}^{\rm D3})
\nonumber\\&
=\cdots(\text{identical terms})\cdots\nonumber\\
&\quad+3 \left(9 u^5 \chi _2^J+5 \chi_6 u^2-5 \chi_4 u+\chi_3  \left(23 u^5-9 u^2-\frac{2}{u}\right)\right.\nonumber\\
&\hspace{2cm}\left.+\chi_1 \left(-23 u^4+9 u+\frac{1}{u^2}\right)+184 u^{8}-53 u^5-26 u^2- \frac{8}{u}-\frac{2}{u^4}\right)q^8\nonumber\\
   &\quad+\mathcal{O}(q^{\frac{17}{2}}).
\end{align}
The second and the third relations in (\ref{eq114}) are related by the Weyl reflection.
The second relation is confirmed for $N=3$ by comparing the following results.
\begin{align}
\frac{{\cal I}^{\rm gauge}_{\bm{b}_e-\bm{b}_{w_Y}}}{{\cal I}^{\rm gauge}_0}
&=\frac{v^3}{u^3}q^3+\left(\frac{1}{u^2}-\frac{v^3}{u^2}\right)q^4+\frac{v^3 }{u^3}\chi _1^Jq^{\frac{9}{2}}+\left(\frac{1}{uv^3}-\frac{v^3}{u}\right)q^5
\nonumber\\
&\quad+\left(\frac{v^3 }{u^3}\chi
   _2^J-\frac{v^3}{u^3}-\frac{1}{u^3}+\frac{1}{v^6}-1\right)q^6+\left(\frac{1}{u}-\frac{v^3}{u}\right)\chi _1^Jq^{\frac{13}{2}}\nonumber\\
   &\quad+ \left(\frac{v^3}{u^2}-\frac{1}{u^2}-\frac{2u}{v^6}+uv^3-\frac{2u}{v^3}-3u\right)q^7
   +\mathcal{O}(q^{\frac{15}{2}}),\\
{\cal I}_{Y,0}^{\rm D3}
&=\cdots(\text{identical terms})\cdots\nonumber\\
&\quad+ \left( \frac{u}{v^9}-u+\frac{v^3}{u^2}-\frac{1}{u^2}\right)q^7+\mathcal{O}(q^{\frac{15}{2}}).
\end{align}

%%%%%%%%%%%%%%%%%%%%%%%%%%%%%%%%%%%%%%%%%%%%%%
\subsection{$\CC^3/\ZZ_4$}
The first relation in (\ref{z44expected}) is confirmed for $N=3$ by comparing the following results.
\begin{align}
{\cal I}^{\rm gauge}_0
&=1+2u^2q^2+\left(5u^4-\frac{\chi_2}{u}+\chi_4\right)q^4+2u\left(5u^5- \chi_2+u \chi_4\right)q^6
\nonumber\\
&\quad-4u^4q^7+\mathcal{O}(q^{\frac{15}{2}}),\\
\mathcal{I}^{\mathrm{grav}} {\cal I}_X^{\rm tensor}
&=\cdots(\text{identical terms})\cdots+0q^7+\mathcal{O}(q^8),
\end{align}
The second relation in (\ref{z44expected}) is confirmed for $N=3$ by comparing the following results.
\begin{align}
&\frac{{\cal I}^{\rm gauge}_{\bm{b}_e-\bm{b}_{w_Y}}}{{\cal I}^{\rm gauge}_0}
\nonumber\\&
=\chi_3 q^3
+\chi _1^J \left(\chi_3 -\frac{\chi_1}{u}\right)q^{\frac{9}{2}}
+ u\left(\chi_1-\chi_3 u\right)q^5
\nonumber\\
&\quad+ \left(\chi_3 \left(\chi _2^J-1\right)-\chi_1
   \left(\frac{\chi _2^J}{u}+\frac{1}{u}\right)\right)q^6
+ u\chi_1  \chi _1^Jq^{\frac{13}{2}}-\left(\chi_3
   u^4+\frac{\chi_1}{u^3}\right)q^7
\nonumber\\
&\quad+ \left(\chi_1 \left(\frac{2 \chi _1^J}{u}-\frac{\chi
   _3^J}{u}\right)-\chi_3 \left(\chi _1^J-\chi _3^J\right)\right)q^{\frac{15}{2}}\nonumber\\
   &\quad+ u \left(u\chi_3 -4
   \chi_1\right)q^8+\chi _1^J \left(\chi_1 \left(3 u^3-\frac{1}{u^3}\right)-\chi_3
   \left(u^4-\frac{1}{u^2}\right)\right)q^{\frac{17}{2}}\nonumber\\
   &\quad+ \left(\chi_1 \left(\frac{\chi _2^J}{u}-\frac{\chi _4^J}{u}-2
   u^5+\frac{2}{u}\right)-\chi_3 \left(\chi _2^J-\chi _4^J+3\right)\right)q^9\nonumber\\
   &\quad+\left(\chi_1\left(\frac{2}{u^3}-3u^3+\chi_2^Ju^3\right)+\chi_3\left(\frac{1}{u^2}+u^4\right)-\frac{2}{u}\chi_5\right)q^{10}+\mathcal{O}(q^{\frac{21}{2}}),\\
%%%
&{\cal I}_{Y,0}^{\rm D3}+{\cal I}_{Z,0}^{\rm D3}
\nonumber\\&
=\cdots(\text{identical terms})\cdots\nonumber\\
&\quad+\left(\chi_1\left(\frac{2}{u^3}-u^3+u^3\chi_2^J\right)+\frac{\chi_3}{u^2}-u^4\chi_3-\frac{2}{u}\chi_5\right)q^{10}+\mathcal{O}(q^{\frac{21}{2}}).
\end{align}
The third relation in (\ref{z44expected}) is confirmed for $N=3$ by comparing the following results.
\begin{align}
\frac{{\cal I}^{\rm gauge}_{\bm{b}_e-\bm{b}_{w_Y^2}}}{{\cal I}^{\rm gauge}_0}
&= u^3q^3
  +  u^2\chi_2q^4
  + u^3 \chi _1^Jq^{\frac{9}{2}}
  + \left(u\chi_4+\frac{1}{u}\right)q^5
\nonumber\\
 &\quad+ \left(u^3 \left(\chi_2^J-1\right)+\frac{\chi_2}{u^2}+\chi_6\right)q^6
  -\chi_2  \chi _1^Jq^{\frac{13}{2}}
  - u^2\left(\chi_4 u+\chi_2\right)q^7
\nonumber\\
&\quad+\mathcal{O}(q^{\frac{15}{2}}),\\
{\cal I}_{X,0}^{\rm D3}
&=\cdots(\text{identical terms})\cdots
\nonumber\\
&\quad+
   \left(-\chi_2 u^2+\frac{\chi_8}{u}+\chi_4 \left(\frac{1}{u^3}-u^3\right)+\frac{1}{u^5}\right)q^7+\mathcal{O}(q^{\frac{15}{2}}).
\end{align}

%%%%%%%%%%%%%%%%%%%%%%%%%%%%%%%%%%%%%%%%%%%%%%
\subsection{$\CC^3/(\ZZ_2\times\ZZ_2)$}
The first relation in (\ref{z2z2expected}) is confirmed for $N=3$ by comparing the following results.
\begin{align}
\mathcal{I}^{\text{gauge}}_0
&=1+2\left( \chi _{(2,0)}- \chi _{(0,1)}\right)q^2 +\left(4 \chi _{(0,2)}+\chi _{(1,0)}-5 \chi _{(2,1)}+5 \chi _{(4,0)}\right)q^4 \nn\\
&\quad+ \left(-10 \chi _{(0,3)}+10 \chi _{(2,2)}-10\chi _{(4,1)}+10 \chi _{(6,0)}-2\right)q^6\nn\\
&\quad-4 \left(\chi _{(0,2)}-\chi _{(2,1)}+\chi _{(4,0)}\right)q^7 +\mathcal{O}(q^8)
\end{align}
and on the gravity side we have
\begin{align}
{\cal I}^{\rm grav}&
{\cal I}^{\rm tensor}_X
{\cal I}^{\rm tensor}_Y
{\cal I}^{\rm tensor}_Z
=\cdots(\text{identical terms})\cdots+0q^7+\mathcal{O}(q^8)
\end{align}
The second, the third, and the fourth relations in (\ref{z2z2expected}) are related by Weyl reflections among $X$, $Y$ and $Z$.
The second relation in (\ref{z2z2expected}) is confirmed for $N=3$ by comparing the following results.
\begin{align}
\frac{{\cal I}^{\rm gauge}_{\bm{b}_e-\bm{b}_{w_X}}}{{\cal I}^{\rm gauge}_0}
&= u^3q^3
+ uq^4
+ u^3 \chi _1^Jq^{\frac{9}{2}}
+\left(-\frac{u^3}{v^2}-u v^2+\frac{1}{u}\right)q^5
\nonumber\\&\quad
+ \left(u^3 \chi _2^J-u^3+\frac{1}{u^3}\right)q^6
+\left(-\frac{u^3}{v^4}-\frac{v^4}{u}-u\right)q^7+\mathcal{O}(q^{\frac{15}{2}})
\end{align}
\begin{align}
{\cal I}^{\rm D3}_{X,0}
&=\cdots(\text{identical terms})\cdots+\left(\frac{1}{u^5}-\frac{u^3}{v^4}-\frac{v^4}{u}-u\right)q^7+\mathcal{O}(q^{\frac{15}{2}})
\end{align}

%%%%%%%%%%%%%%%%%%%%%%%%%%%%%%%%%%%%%%%%%%%%%%

%%%%%%%%%%%%%%%%%%%%%%%%%%%%%%%%%%%%%%%%%%%%%%%%%%%%%%%%%
\end{document}